\documentclass[english]{article}
\usepackage[T1]{fontenc}
\usepackage[latin9]{inputenc}
\usepackage{array}
\usepackage{booktabs}
\usepackage{units}
\usepackage{multirow}
\usepackage{amsmath}
\usepackage{amsthm}
\usepackage{amssymb}
\usepackage{graphicx}

\makeatletter

\providecommand{\tabularnewline}{\\}

\numberwithin{equation}{section}
\numberwithin{figure}{section}

\usepackage{babel}
\usepackage{fullpage}
\usepackage{authblk}
\usepackage{hyperref}
\usepackage{placeins}
\usepackage{hyphenat}
\usepackage{listings}
\usepackage{fancyvrb}
\usepackage[hypcap]{caption}

\usepackage[title,titletoc]{appendix}

\title{Taxation of a GMWB Variable Annuity in a Stochastic Interest Rate Model}
\author{ \textsc{Andrea Molent}\thanks{Dipartimento di Scienze Economiche e Statistiche, Universit\`a degli Studi di Udine, Italy - \texttt{andrea.molent@uniud.it}}}
\date{}

\makeatother

\usepackage{babel}
\begin{document}
\maketitle

\begin{flushleft}
\rule{1\columnwidth}{1pt}
\par\end{flushleft}

\begin{flushleft}
\textbf{\large{}Abstract}
\par\end{flushleft}{\large \par}

Modeling taxation of Variable Annuities has been frequently neglected
but accounting for it can significantly improve the explanation of
the withdrawal dynamics and lead to a better modeling of the financial
cost of these insurance products. The importance of including a model
for taxation has first been observed by Moenig and Bauer \cite{moenig2016}
while considering a GMWB Variable Annuity. In particular, they consider
the simple Black-Scholes dynamics to describe the underlying security.
Nevertheless, GMWB are long term products and thus accounting for
stochastic interest rate has relevant effects on both the financial
evaluation and the policy holder behavior, as observed by Goudenège
et al. \cite{goudenege2018}. In this paper we investigate the outcomes
of these two elements together on GMWB evaluation. To this aim, we
develop a numerical framework which allows one to efficiently compute
the fair value of a policy. Numerical results show that accounting
for both taxation and stochastic interest rate has a determinant impact
on the withdrawal strategy and on the cost of GMWB contracts. In addition,
it can explain why these products are so popular with people looking
for a protected form of investment for retirement.

\vspace{2mm}

\noindent \emph{\large{}Keywords}: Variable Annuities, taxation,
stochastic interest rate, optimal withdrawal, tree method.

\noindent\rule{1\columnwidth}{1pt}

\newpage

\section{Introduction}

Variable Annuities are tax deferred investment contracts with insurance
coverage. The market for such products has been steadily growing in
the past years all around the world and 2019 has set best sales year
since 2008 in Unites States. According to the Secure Retirement Institute
\cite{SRI2019}, the Variable Annuity sales in 2019 amounted to over
\$100 billions, which represents almost half of the total annuity
sales. In this paper we focus on a particular type of Variable Annuity,
called Guaranteed Minimum Withdrawal Benefit (GMWB) which promises
to return the entire initial investment by means of cash withdrawals
during the policy life, plus a final payment amounting to the remaining
account value at the contract maturity. Usually, the policy holder
(hereinafter PH) pays the whole premium as a lumpsum and he is entitled
to withdraw at each contract anniversary a variable amount, with a
minimum guaranteed. Thanks to the guarantee included in the policy,
the PH can withdraw money from his account even if it has run out.
Moreover, if the PH death occurs before the contract maturity, then
his heirs receive the remaining account value as a lumpsum payout.
The premium paid at contract inception determines the risky account,
which changes over time according to a financial index (usually a
fund) but it is also reduced due to the fees applied by the insurer
and by withdrawals made by the PH.

In order to manage GMWB contracts, insurers usually employ hedging
techniques which rely on the computation of the fair prices of the
policies in a risk neutral probability framework. In addition, the
hedging costs are offset by deducting a proportional fee from the
risky asset account. Moreover, the mortality risk is hedged by using
the law of large numbers (see Bernard and Kwak \cite{bernard2016}
and Lin et al. \cite{lin2016} for an explanation of move-based and
semi-static hedging of Variable Annuities). Price and Greeks calculation
usually relies on numerical computations, which are based on a convenient
model of the product, of the financial market, and nonetheless of
the behavior of the PH. In fact, since the PH can choose (within certain
limits established by the contract) the amount to be withdrawn, he
can decisively drive the total payoff of the contract. Anyway, ordinary
techniques for pricing American and Bermudan options lead to prices
which differ significantly from market observations (Moenig and Bauer
\cite{moenig2016}). A possible explanation to the theoretical-empirical
price gap, can be found in a correct model for the dynamics of taxation
that the customer must face. In this regard, Moenig and Bauer \cite{moenig2016}
propose to model taxation imposed to the PH and to consider a subjective
valuation of the contract. Specifically, they show that when accounting
for taxation, PH withdraws less frequently than without taxes and
by employing ordinary pricing techniques, one can obtain prices which
are in line with empirical observations. Moreover, Moenig and Zhu
\cite{moenig2018} observe that the preferential tax treatment has
been one of the key factors that have made Variable Annuities such
a popular instrument and thus correctly modeling taxation can improve
the explanation of the still unclear mechanisms about these products.
We stress out that the investigations in \cite{moenig2016} and \cite{moenig2018}
have been performed by assuming the Black-Scholes model for the underlying
fund.

Interest rates is another relevant factor in Variable Annuities evaluation.
As observed by Goudenège et al. \cite{goudenege2018}, since GMWB
contracts have long maturities that could last almost 25 years, the
Black-Scholes model seems to be unsuitable for such a long time interval
as it assumes constant interest rate and volatility. Several authors
have investigated the possibility of evaluating GMWB contracts while
considering a stochastic interest rate. For example, Peng et al. \cite{peng2012}
develop an analytic approximation of the fair value of the GMWB under
the Vasicek stochastic interest rate model. Donnelly et al. \cite{donnelly2012}
consider pricing and Greeks calculation through an Alternating Direction
Implicit method in the advanced Heston-Hull-White model. Dai et al.
\cite{dai2015} develop a tree based model to include both stochastic
interest rate and mortality in their evaluation framework. Gudkov
et al. \cite{gudkov2017} employ the operator splitting method to
price GMWB products under stochastic interest rate, volatility and
mortality. Shevchenko and Luo \cite{shevchenko2017} employ high order
Gauss-Hermite quadrature to evaluate the GMWB contract under the Vasicek
interest rate model. Recently, Goudenège et al. \cite{goudenege2019}
exploit a hybrid tree-PDE method together with Machine Learning techniques
to efficiently evaluate the GMWB contract in a model that considers
both stochastic interest rate and stochastic volatility. More generally,
as far as pricing of Variable Annuities in a stochastic interest rate
framework is considered, it is worth mentioning the work of Bacinello
and Zoccolan \cite{bacinello2019} that develops a Monte Carlo flexible
approach to study the impact of threshold fee on the optimal surrender
strategy about a product including accumulation and death guaranteed
benefits under a model which considers stochastic interest rate, volatility
and mortality. We also mention Goudenège et al. \cite{goudenege2016},
who employ the hybrid Tree-PDE method to evaluate a GLWB contract
under stochastic interest rate.

In this paper we present an investigation about GMWB pricing and PH
behavior when both tax treatment and stochastic interest rate are
considered. In particular, following Moenig and Bauer \cite{moenig2016}
and Moenig and Zhu \cite{moenig2018}, we model taxation of GMWB through
a constant marginal income tax rate on all policy earnings and a constant
marginal tax rate on capital gains from investments outside of the
policy. Moreover, we also include a premium based model for taxation
of the insurer, which was neglected in previous researches. Because
of taxation, the evaluation of the contract is not straightforward,
so we exploit the same subjective risk-neutral valuation methodology
employed in \cite{moenig2016}. In particular, in this framework,
the value of a given post-tax cash flow is the amount necessary to
set up a pre-tax portfolio that replicates the considered cash flow.
This causes the insurer and the PH to evaluate the policy differently
and we investigate both the two perspectives. As far as the stochastic
interest rate is concerned, we consider the Hull-White model (Hull
\cite{hull1994}), which is often employed by both academics and practitioners
for its easiness of calibration and simple probability distribution.
This model has already been employed in other research works concerning
GMWB Variable Annuities (e.g. \cite{donnelly2012}, \cite{dai2015},
\cite{goudenege2018} and \cite{goudenege2019}). We stress out that
considering both taxation and stochastic interest rate is a challenging
task because of the computational effort required to consider many
factors together. In particular, evaluating a GMWB policy in the considered
model is a four (plus time) dimensional problem, which means a high
computational cost in terms of both computing time and working memory
required. Moreover, the evaluation of a policy through the subjective
risk-neutral valuation methodology requires the resolution of many
fixed point problems, and this increases even more the computational
cost. Finally, we assume the PH to employ an optimal withdrawal strategy,
which implies the numerical resolution of a dynamic control problem.
In order to manage such a computational effort, we use a backward
dynamic approach that exploits a tree approach to compute the fair
contract price. In particular, we employ a trinomial tree to approximate
the stochastic interest rate process through a Markov chain, which
represents an efficient numerical solution already used by Goudenège
et al. \cite{goudenege2019}. It is worth noting that tree methods
have already been used to study the GMWB contract. In this regard,
we mention the works of Costabile \cite{costabile2017} and of Costabile
et al. \cite{costabile2020} that employ a trinomial tree to evaluate
a GMWB policy and to investigate the PH decisions while including
exogenous factors in the model.

In order to test our approach, we perform some numerical experiments.
Specifically, we study how the evaluation of the policy varies according
to the insurer and to the PH perspectives and how the withdrawal strategy
is modified, by including or not including taxation and by changing
the parameters of the interest rate and the fund. Numerical results
show many interesting findings. First of all, if taxation is considered,
the fair value of the policy for the PH is higher than the fair value
for the insurer. This means that the PH attributes a higher price
to the policy than the insurer does, so buying and selling the contract
can be a good deal for both of them. Secondly, we observe that taxation
and interest rate modeling have a significant impact on the withdrawal
strategies of the PH.

To the best of our knowledge, this is the first analysis about GMWB
pricing and withdrawal strategy which accounts for both taxation and
stochastic interest rate. Our research could be useful both for the
qualitative observations obtained and for the numerical solutions
adopted.

The reminder of the paper is organized as follows. Section 2 introduces
the stochastic model for the underlying and the interest rate processes.
Section 3 describes the GMWB contract and the taxation model. Section
4 presents pricing assumptions. Section 5 describes the pricing method
and the technical measures. Section 6 shows numerical results on various
examples. Finally, Section 7 draws the conclusions. 

\section{The Stochastic Model \label{Sec2}}

In order to define the notation used throughout the rest of the paper,
let us introduce the Black-Scholes Hull-White model. The Hull-White
model \cite{hull1994} is one of historically most important interest
rate models, which is nowadays often used for option pricing purposes.
In particular, the existence of closed formulas for the price of bonds,
caplets and swaptions is one of the important advantages of this model.
Furthermore, it is capable of generating negative interest rates,
actually observed in the markets in recent years. We report the dynamics
of the Black-Scholes Hull-White model, which combines the dynamics
of the interest rate with the dynamics of the underlying:
\begin{equation}
\begin{cases}
dS_{t} & =r_{t}S_{t}dt+\sigma S_{t}dZ_{t}^{S}\\
dr_{t} & =k\left(\theta_{t}-r_{t}\right)dt+\omega dZ_{t}^{r},
\end{cases}\label{eq:model}
\end{equation}
where $Z^{S}$ and $Z^{r}$ are Brownian motions with $d\left\langle Z_{t}^{S},Z_{t}^{r}\right\rangle =\rho dt$.
Moreover $\sigma,$ $k$ and $\omega$ are positive values and the
initial values $S_{0}>0$ and $r_{0}$ are given. Furthermore, $\theta_{t}$
is a deterministic function which is completely determined by the
market values of the zero-coupon bonds by calibration (see Brigo and
Mercurio \cite{brigo2007}) so that the theoretical prices of the
zero-coupon bonds match exactly the market prices. 

Let $P^{M}\left(0,T\right)$ denote the market price of the zero-coupon
bonds at time $0$ for the maturity $T$. The market instantaneous
forward interest rate is then defined by
\begin{equation}
f^{M}\left(0,T\right)=-\frac{\partial\ln P^{M}\left(0,T\right)}{\partial T}.
\end{equation}
It is well known that the (short) interest rate process $r$ can be
written as 
\begin{equation}
r_{t}=Y_{t}+\beta\left(t\right),\label{eq:rY}
\end{equation}
 where $Y$  is a stochastic process whose dynamics is given by 
\begin{equation}
dY_{t}=-kY_{t}dt+\omega dZ_{t}^{r},\ Y_{0}=0,
\end{equation}
and $\beta\left(t\right)$ is a real valued function with
\begin{equation}
\beta\left(t\right)=f^{M}\left(0,t\right)+\frac{\omega^{2}}{2k^{2}}\left(1-\exp\left(-kt\right)\right)^{2}.
\end{equation}
Then, the Black-Scholes Hull-White model can be described by the following
relations:
\begin{equation}
\begin{cases}
dS_{t}=r_{t}S_{t}dt+\sigma S_{t}dZ_{t}^{S} & S_{0}=\bar{S}_{0},\\
dY_{t}=-kY_{t}dt+\omega dZ_{t}^{r} & Y_{0}=0,\\
r_{t}=Y_{t}+\beta\left(t\right).
\end{cases}\label{eq:HW}
\end{equation}
The \emph{flat curve} case is a particular case for the market price
of a zero-coupon bonds: in this specific case, the price at time $t$
of a zero-coupon bond with maturity $\bar{t}$ is given by 
\begin{equation}
P^{M}\left(t,\bar{t}\right)=e^{-r_{0}\left(\bar{t}-t\right)},
\end{equation}
 and the function $\beta$ is given by 
\begin{equation}
\beta\left(t\right)=r_{0}+\frac{\omega^{2}}{2k^{2}}\left(1-\exp\left(-kt\right)\right)^{2}.
\end{equation}
We stress out that assuming a flat curve for the price of bonds is
not essential for the development of our model, but it simplifies
the numerical settings.

\section{Modeling the contract}

\subsection{Modeling taxation}

In order to model taxation, we follow the same approach proposed by
Moenig and Bauer \cite{moenig2016}, which in turn is a simplified
version of the model currently in force in the Unites States. 

As far as the PH is concerned, taxes are due on future investment
gains and not on the invested amount. In particular, we assume a constant
marginal income tax rate $\tau$ to be applied on all policy earnings
and a constant marginal tax rate $\kappa$ to be applied to the capital
gains from investments outside of the policy. This means that if the
PH sets up a portfolio that replicates the after tax policy cash flows,
then $\kappa$ is the tax rate applied on gains of such a portfolio.
On the contrary, $\tau$ is the tax rate applied on all gains caused
by PH's withdrawals form the policy. In particular, earnings are withdrawn
before the initial premium, following last-in first-out approach. 

In order to complete tax modeling we have to consider taxation concerning
the insurer, which is usually of two types: premium taxation and net
income taxation (see Skipper \cite{skipper2001}). Determining life
insurer profit is a challenge because of the difference in timing
between premium payments and claim payments, so premium taxes are
the most common. Furthermore, as far as Unites States life insurance
system is concerned, the insurance companies can elect to be taxed
based on either premiums or net income (see Nissim \cite{nissim2010}).
For for sake of simplicity, we assume premium based taxation, that
is the insurer pays a certain percentage of the gross premium $GP$
as taxes. So, the tax due by the insurer is thus $\chi\cdot GP$,
where $\chi$ is the premium tax rate. Such a rate usually varies
between $0.5\%$ and $3\%$ (see Moran \cite{winsconsin2017}). Obviously
the insurer has to recover this tax cost, therefore we assume that
such an amount is applied indirectly to the customer as an entry cost,
which reduces the gross premium and determines the net premium $P$,
given by $P=GP\cdot\left(1-\chi\right)$.

\subsection{The GMWB contract}

We study here a simple version of the GMWB contract which was first
investigated by Moenig and Bauer \cite{moenig2016}. We consider an
$x$-year old individual that purchased a GMWB policy with a finite
integer maturity $T$ against the payment of a single gross premium
$GP$. Then entry expenses are deducted from the gross premium and
the net premium $P$ is credited to the policy\textquoteright s account.
There are three variables which determine the state of a policy at
time $t$, namely the account value $X_{t}$, the benefit base $G_{t}$
and the tax base $H_{t}$ whose values at time $t=0$ are equal to
the policy net premium, that is
\begin{equation}
X_{0}=G_{0}=H_{0}=P.\label{eq:initial}
\end{equation}
In particular, the account value $X$ represents the risky account
of the policy, which changes as if it were invested in a market fund,
aside from being reduced by withdrawals and management costs. The
benefit base $G$ represents the guarantee inherent in the policy
as it regulates the maximum withdrawal that the PH can make, while
the tax base $H$ represents the amount that may still be withdrawn
from the policy free of tax.

Let $t_{i}$ denote the time of the $i-th$ contract anniversary,
i.e. $t_{i}=i$. The variables $G_{t}$ and $H_{t}$ do not change
during the time between two consecutive anniversaries, that is for
$t\in\left]t_{i-1},t_{i}\right[$, while $X_{t}$ varies according
to an underlying investment fund changes. This fund is usually chosen
by the customer from a list proposed by the insurer. Specifically,
let us term $S_{t}$ the value of the underlying fund, which evolves
according to (\ref{eq:model}). Then, for $t\in\left]t_{i-1},t_{i}\right[$
, $X_{t}$ follows the same dynamics of $S_{t}$ with the exception
that fees are subtracted continuously, that is

\begin{equation}
dX_{t}=\frac{X_{t}}{S_{t}}dS_{t}-\varphi X_{t}dt.\label{eq:fees_cont}
\end{equation}
The variable $\varphi$ in (\ref{eq:fees_cont}) is the (constant)
fee rate and it controls the fees withdrawn by the account value. 

At each anniversary time $t_{i}$, the continuation of the policy
is determined according to the survival of the PH during the last
year of the contract. In order to describe the policy revaluation
mechanisms, let us denote with $X_{t_{i}}^{-}$and $X_{t_{i}}^{+}$
the account values just before and after any cash-flow at time $t_{i}$
(we use the same notation for $G_{t_{i}}$ and $H_{t_{i}}$). If the
PH has passed away during the previous year, then his heirs receive
the death benefit $b_{i}$, which is paid at time $t_{i}$ and it
is given by the residual account value net of taxation, that is
\begin{equation}
b_{i}=X_{t_{i}}^{-}-\tau\left(X_{t_{i}}^{-}-H_{t_{i}}^{-}\right)_{+},\label{eq:db}
\end{equation}
where $\tau$ is the income tax rate and $\left(x\right)_{+}=\max\left(x,0\right)$.
After the payment of the death benefit, the contract ends and it has
no residual value. On the contrary, if the PH has not passed away,
then he is entitled to withdraw an amount $w_{i}$ within some limits.
According to the contract, the withdrawal amount $w_{i}$ selected
by the PH must satisfy the following relation:
\begin{equation}
0\leq w_{i}\leq\max\left\{ X_{t_{i}}^{-},\min\left\{ g^{W},G_{t_{i}}^{-}\right\} \right\} ,
\end{equation}
where $g^{W}$ is a positive constant value called the annual guaranteed
amount and it is stated in the contract. In particular, if $g^{W}=\nicefrac{P}{T}$
then the PH is entitled to withdraw at each contract anniversary exactly
an amount equal to $g^{W}$ throughout the duration of the contract.
After the withdrawal has been performed, the new account value is
given by
\begin{equation}
X_{t_{i}}^{+}=\left(X_{t_{i}}^{-}-w_{i}\right)_{+},\label{eq:X+}
\end{equation}
while the new benefit base and tax base are given by
\begin{equation}
G_{t_{i+1}}^{-}=G_{t_{i}}^{+}=\begin{cases}
\left(G_{t_{i}}^{-}-w_{i}\right)_{+} & \text{, if }w_{i}\leq g^{W}\\
\left(\min\left\{ G_{t_{i}}^{-}-w_{i},G_{t_{i}}^{-}\cdot\frac{X_{t_{i}}^{+}}{X_{t_{i}}^{-}}\right\} \right)_{+} & \text{, if }w_{i}>g^{W}
\end{cases}\label{eq:G+}
\end{equation}
and 
\begin{equation}
H_{t_{i+1}}^{-}=H_{t_{i}}^{+}=H_{t_{i}}^{-}-\left(w_{i}-\left(X_{t_{i}}^{-}-H_{t_{i}}^{-}\right)_{+}\right)_{+}\label{H+}
\end{equation}
respectively. 

The PH does not receive the whole amount withdrawn $w_{i}$ because
some fees and tax may be applied. Specifically, the PH receives the
withdrawn amount reduced by the fees due to the insurer for withdrawing
more than the guaranteed amount $g^{W}$ and also reduced by a penalty
for early withdrawals and by the taxation. Specifically, the net amount
he receives is given by
\begin{equation}
w_{i}-fee_{i}-pen_{i}-tax_{i}
\end{equation}
being $fee_{i}$ the cost for withdrawing an amount exceeding $\min\left\{ g^{W},G_{t_{i}}\right\} $,
$pen_{i}$ an early withdrawal penalty for any withdrawal before the
age of 59.5 years and $tax_{i}$ the income taxes associated with
the withdrawal. In particular,
\begin{equation}
fee_{i}=s_{i}\cdot\left(w_{i}-\min\left\{ g^{W},G_{t_{i}}^{-}\right\} \right)_{+},\label{eq:fee}
\end{equation}
\begin{equation}
pen_{i}=s^{g}\cdot\left(w_{t}-fee_{i}\right)\cdot1_{\left\{ x+t_{i}<59.5\right\} },\label{eq:pen}
\end{equation}
and
\begin{equation}
tax_{i}=\tau\cdot\min\left\{ w_{i}-fee_{i}-pen_{i},\left(X_{t_{i}}^{-}-H_{t_{i}}^{-}\right)_{+}\right\} .\label{eq:tax}
\end{equation}
The coefficient $s_{i}$ in (\ref{eq:fee}) is a non-negative coefficient
called surrender charge, which usually decreases with time and it
is zero within the term of the contract. Moreover, $s^{g}$ in (\ref{eq:pen})
is another non-negative coefficient that determines the penalty for
an early withdrawal. In particular, since these contracts are usually
employed as a supplement to the retirement pension, we assume that
when the contract maturity is achieved, the PH must be older than
$59.5$ years, so penalty is not applied at last withdrawal at time
$T$. 

Finally, after the last withdrawal has been made at time $T$, the
alive PH receives the remaining account value net of taxes, that is
\begin{equation}
X_{T}^{+}-\tau\left(X_{T}^{+}-H_{T}^{+}\right)_{+},
\end{equation}
and the contract ends.

\section{Pricing assumptions}

In this Section we present the pricing framework. First of all, we
observe that asset pricing under taxation is not direct since, as
observed by Ross \cite{ross1987}, taxation leads to the loss of uniqueness
of prices. In fact, the valuation of a specific cash flow depends
on the personal endowment and tax rates. Following the same approach
of Moenig and Bauer \cite{moenig2016}, we define the value of a given
post-tax cash flow as the amount of money that a specific agent needs
to create a financial portfolio (made of stocks and bonds) that, after
taxation, replicates the considered cash flow. Clearly, such a value
depends on the specific taxation applied to the agent and this can
be significantly different between the customer and the insurer. In
addition, taxation may be different in relation to the financial instrument
considered: a lighter taxation is usually applied to insurance products
(such as VA policies) and a heavier taxation for financial products
(such as the securities in a replicating portfolio).

First of all, we present how to evaluate the GMWB contract assuming
PH's subjective valuation and then we present the same while assuming
insurer's subjective valuation. The main differences in the two perspectives
are due to taxation and to control on withdrawals. As far as taxation
is considered, the PH has to pay taxes on both policy earnings and
capital gains outside the policy. On the contrary, the taxation applied
to the insurer is much simpler: a percentage of the gross premium.
As far as withdrawals are concerned, the PH selects optimal withdrawals
in order to maximize the expected value of its assets, net of taxation:
if taxation is applied, such a value is not equal to insurer's liability.
Thus, the amount withdrawn by the PH is optimal for him, but it could
be different from the worst amount computed considering the insurer's
point of view, that is the amount that maximizes insurer's liability
to the PH. This means that the PH withdraws money trying to maximize
his economic return, rather than trying to maximize the outputs of
the insurer: since these two strategies do not coincide, the costs
for the insurer are lower than the worst withdrawing case.

Finally, we underline that the considered framework captures an interesting
feature of insurance products. Taxation makes GMWB policies particularly
attractive to customers: although taxes are applied on the earnings
of the policy, the tax regime is particularly favorable for this type
of product and therefore it is more convenient for the customer buying
the policy rather than reproducing it through a replicating portfolio.

In the next Subsections, we show how to compute the initial contract
value according to the PH and to the insurer's subjective valuation.
We stress out that in both cases we compute the cost of the replicating
portfolio under the same risk neutral measure $\mathbb{Q}$ for the
Black-Scholes Hull-White model (see Brigo and Mercurio \cite{brigo2007}).

\subsection{Policyholder\textquoteright s subjective valuation}

Following Moenig and Bauer \cite{moenig2016}, we consider the PH\textquoteright s
subjective valuation of the contract. This means we compute the amount
of money that a PH needs to set a hypothetical replicating portfolio,
which replicates the post-tax policy cash-flows. Specifically, let
$\mathcal{V}\left(t,r,X,G,H\right)$ denote the fair value according
to an alive PH of a GMWB contract at time $t$, being $r$ the interest
rate, $X$ the account value, $G$ the guarantee base and $H$ the
tax base respectively. Specifically, following the same approach of
Moenig and Bauer \cite{moenig2016}, $\mathcal{V}$ represents the
average option value across the many policies sold to the customers
that are still alive at time $t$. 

Finally, in order to compute PH's subjective value of the contract
at time $t=0$, we proceed backward in time, starting from contract's
maturity at time $T$ and by taking into account the changes that
occur to the policy status parameters.

\subsubsection{Value function at a contract anniversary}

First of all, let us denote with $\mathcal{V}^{+}\left(T,r_{T},X_{T}^{+},G_{T}^{+},H_{T}^{+}\right)$
the policy value at maturity, after the last withdrawal is performed.
Such an amount is given by the final payoff, that is
\begin{equation}
\mathcal{V}^{+}\left(T,r_{T},X_{T}^{+},G_{T}^{+},H_{T}^{+}\right)=X_{T}^{+}-\tau\left(X_{T}^{+}-H_{T}^{+}\right)_{+}.\label{eq:Terminal}
\end{equation}
Now, let us focus on the $i$-th contract anniversary, at time $t_{i}$.
Since we are assuming that the PH is alive, then he is entitled to
perform a withdrawal from his account. Let $\mathcal{V}^{-}\left(t_{i},r_{t_{i}},X_{t_{i}}^{-},G_{t_{i}}^{-},H_{t_{i}}^{-}\right)$
and $\mathcal{V}^{+}\left(t_{i},r_{t_{i}},X_{t_{i}}^{+},G_{t_{i}}^{+},H_{t_{i}}^{+}\right)$
represent the values of the policy just before and after the PH has
withdraw money respectively. In particular, $r_{t_{i}},X_{t_{i}}^{-},G_{t_{i}}^{-},H_{t_{i}}^{-}$
are the state parameters before withdrawing at time $t_{i}$, while
$r_{t_{i}},X_{t_{i}}^{+},G_{t_{i}}^{+},H_{t_{i}}^{+}$ are the state
parameters after withdrawing at time $t_{i}$. Please, observe that
there is no need to distinguish between the value of the interest
rate before and after the withdrawal, because such a value is not
modified by the withdrawal, so we simply write $r_{t_{i}}$ in both
cases. We can write the relation between the two policy values in
the general form
\begin{equation}
\mathcal{V}^{-}\left(t_{i},r_{t_{i}},X_{t_{i}}^{-},G_{t_{i}}^{-},H_{t_{i}}^{-}\right)=\mathcal{V}^{+}\left(t_{i},r_{t_{i}},X_{t_{i}}^{+}\left(w_{i}\right),G_{t_{i}}^{+}\left(w_{i}\right),H_{t_{i}}^{+}\left(w_{i}\right)\right)+\left(w_{i}-fee_{i}\left(w_{i}\right)-pen_{i}\left(w_{i}\right)-tax_{i}\left(w_{i}\right)\right),\label{eq:V_t_minus}
\end{equation}
where we underline the dependence of many variables on the withdrawal
$w_{i}$ by denoting them as a function of $w_{i}$. In particular,
equations (\ref{eq:X+}), (\ref{eq:G+}), (\ref{H+}), (\ref{eq:fee}),
(\ref{eq:pen}) and (\ref{eq:tax}) express the dependence of $X_{t_{i}}^{+},G_{t_{i}}^{+},H_{t_{i}}^{+},fee_{i},pen_{i}$
and $tax_{i}$ on $w_{i}$ respectively. The PH might adopt a static
withdrawal strategy, which means he withdraw an amount $w_{i}$ equal
to $g^{W}$, regardless of the value taken from the policy state parameters.
Such a strategy is easy to be implemented, but may be not the optimal
one for him. We rather assume that the PH selects the amount $w_{i}$
in order to maximize the expected value of his assets -- contract
plus net withdrawal--, that is 
\begin{equation}
w_{i}=\underset{w\in\left[0,W_{max}\right]}{\mbox{argmax}}\ \mathcal{V}^{+}\left(t_{i},r_{t_{i}},X_{t_{i}}^{+}\left(w\right),G_{t_{i}}^{+}\left(w\right),H_{t_{i}}^{+}\left(w\right)\right)+\left(w-fee_{i}\left(w\right)-pen_{i}\left(w\right)-tax_{i}\left(w\right)\right),\label{eq:argmax}
\end{equation}
where 
\begin{equation}
W_{max}=\max\left\{ X_{t_{i}}^{-},\min\left\{ g^{W},G_{t_{i}}^{-}\right\} \right\} \label{eq:I}
\end{equation}
is the maximum withdrawal allowed by the contract. We observe that,
at maturity, the optimization problem (\ref{eq:argmax}) can be easily
solved as the continuation value after the payment is given by the
final payoff, which has a closed formulation. Specifically, one can
prove that the optimal withdrawal in this particular case is given
by
\begin{equation}
w_{T}=\min\left\{ g^{W},G_{T}^{-}\right\} .\label{eq:w_finale}
\end{equation}
Moreover, by using equations (\ref{eq:Terminal}), (\ref{eq:V_t_minus})
and (\ref{eq:w_finale}), one can obtain the following expression:
\begin{multline}
\mathcal{V}^{-}\left(T,r_{T},X_{T}^{-},G_{T}^{-},H_{T}^{-}\right)=\max\left\{ X_{T}^{-},w_{T}\right\} -\tau\min\left\{ w_{T},\left(X_{T}^{-}-H_{T}^{-}\right)_{+}\right\} \\
-\tau\left(\left(w_{T}-\left(X_{T}^{-}-H_{T}^{-}\right)_{+}\right)_{+}+\left(X_{T}^{-}-w_{T}\right)_{+}-H_{T}^{-}\right)_{+}.\label{eq:Final_value}
\end{multline}
In general, when considering the optimal withdrawal at time $t_{i}$,
there is no closed formula as for the last anniversary $T$. In the
general case, the optimal withdrawal $w_{i}$ must be approximated
by a numerical procedure.

\subsubsection{Dynamics of the value function between two anniversaries}

During the time between two contract anniversaries $t_{i}$ and $t_{i+1}$,
the variables $G$ and $H$ do not change. Changes of the policy value
are solely due to the passage of time and to the changes of the account
value $X$ and of the interest rate $r$. Following Moenig and Bauer
\cite{moenig2016}, the subjective risk-neutral value at time $t_{i}$
of $\mathcal{V}^{+}$ is given via a nonlinear implicit equation:
\begin{multline}
\mathcal{V}^{+}=\mathbb{E}^{\mathbb{Q}}\left[e^{-\int_{t_{i}}^{t_{i+1}}r_{s}ds}\left(q_{x+t_{i}}b_{i+1}+p_{x+t_{i}}\mathcal{V}^{-}\right)\right]\\
+\frac{\kappa}{1-\kappa}\cdot\mathbb{E}^{\mathbb{Q}}\left[e^{-\int_{t_{i}}^{t_{i+1}}r_{s}ds}\left(q_{x+t_{i}}b_{i+1}+p_{x+t_{i}}\mathcal{V}^{-}-\mathcal{V}^{+}\right)_{+}\right],\label{eq:NIE}
\end{multline}
where $\mathcal{V}^{+}$ stands for $\mathcal{V}^{+}\left(t_{i},r_{t_{i}},X_{t_{i}}^{+},G_{t_{i}}^{+},H_{t_{i}}^{+}\right)$
and $\mathcal{V}^{-}$ stands for $\mathcal{V}^{-}\left(t_{i+1},r_{t_{i+1}},X_{t_{i+1}}^{-},G_{t_{i+1}}^{-},H_{t_{i+1}}^{-}\right)$.
Furthermore, $b_{i+1}$ is the death benefit that may be paid at time
$t_{i+1}$ in case of death and it is computed according to (\ref{eq:db}).
Moreover, $q_{x+t_{i}}$ is the probability that the alive PH, aged
exactly $x+t_{i}$ at time $t_{i}$, will die in one year, while $p_{x+t_{i+1}}$
is the probability that he will survive at least one more year. We
stress out that the use of death and survival probabilities is possible
if a large number of contract holders is assumed: in this case, mortality
risk is diversifiable.

\subsection{Insurer\textquoteright s subjective valuation}

Let $\mathcal{U}\left(t,r,X,G,H\right)$ denote the fair value of
the GMWB contract but according to insurer's subjective value, that
is the amount of money that the insurer needs to set a replicating
portfolio. The valuation according to the insurer differs from the
valuation according to the PH for some reasons. First of all, the
taxation applied to the insurer only concerns the initial gross premium
and it is not applied to the replicating portfolio. Secondly, the
insurer must shell out an amount gross of taxes, while the PH receives
the net amount. Finally, the insurer has no decision-making power
and undergoes the PH's choices regarding the amount to be withdrawn.
Just as done for the PH's subjective valuation, in order to compute
insurer's subjective value at contract inception, we proceed backward
in time.

\subsubsection{Value function at a contract anniversary}

Let $\mathcal{U}^{+}\left(T,r_{T},X_{T}^{+},G_{T}^{+},H_{T}^{+}\right)$
be the policy value at maturity according to the insurer, after the
last withdrawal is performed. Such an amount is given by the final
payoff before tax, that is the residual account value:
\begin{equation}
\mathcal{U}^{+}\left(T,r_{T},X_{T}^{+},G_{T}^{+},H_{T}^{+}\right)=X_{T}^{+}.\label{eq:Terminal-U}
\end{equation}
Moreover, since the optimal withdrawal $w_{T}$ at time $T$ is given
by (\ref{eq:w_finale}), one can prove the following relation:
\begin{equation}
\mathcal{U}^{-}\left(T,r_{T},X_{T}^{-},G_{T}^{-},H_{T}^{-}\right)=\max\left\{ X_{T}^{-},\min\left\{ g^{W},G_{T}^{-}\right\} \right\} .\label{eq:Final_U}
\end{equation}
Now, let us focus on the $i$-th contract anniversary at time $t_{i}$.
The functions $\mathcal{U}^{-}\left(t_{i},r_{t_{i}},X_{t_{i}}^{-},G_{t_{i}}^{-},H_{t_{i}}^{-}\right)$
and $\mathcal{U}^{+}\left(t_{i},r_{t_{i}},X_{t_{i}}^{+},G_{t_{i}}^{+},H_{t_{i}}^{+}\right)$
represent the value of the contract just before and after the PH has
withdrawn the amount $w_{i}$, which is the solution of problem (\ref{eq:argmax}).
The following relation holds, 
\begin{multline}
\mathcal{U}^{-}\left(t_{i},r_{t_{i}},X_{t_{i}}^{-},G_{t_{i}}^{-},H_{t_{i}}^{-}\right)=\mathcal{U}^{+}\left(t_{i},r_{t_{i}},X_{t_{i}}^{+}\left(w_{i}\right),G_{t_{i}}^{+}\left(w_{i}\right),H_{t_{i}}^{+}\left(w_{i}\right)\right)+\left(w_{i}-fee_{i}\left(w_{i}\right)-pen_{i}\left(w_{i}\right)\right).\label{eq:4.2-IN}
\end{multline}
Equation (\ref{eq:4.2-IN}) is similar to equation (\ref{eq:V_t_minus})
but taxes are not subtracted because the insurer has to pay the amount
before taxation. 

\subsubsection{Dynamics of the value function between two anniversaries}

As opposed to the PH, the insurer pays no taxes on the replicating
portfolio. The subjective risk-neutral value at time $t_{i}^{+}$
of $\mathcal{U}$, is given by the discounted expected future value
of the death benefit plus the value of the policy, that is
\begin{equation}
\mathcal{U}^{+}=\mathbb{E}^{\mathbb{Q}}\left[e^{-\int_{t_{i}}^{t_{i+1}}r_{s}ds}\left(q_{x+t_{i}}b_{i+1}+p_{x+t_{i}}\mathcal{U}^{-}\right)\right],\label{eq:NIE-1}
\end{equation}
 where $\mathcal{U}^{+}$ stands for $\mathcal{U}^{+}\left(t_{i},r_{t_{i}},X_{t_{i}}^{+},G_{t_{i}}^{+},H_{t_{i}}^{+}\right)$
and $\mathcal{U}^{-}$ stands for $\mathcal{U}^{-}\left(t_{i+1},r_{t_{i+1}},X_{t_{i+1}}^{-},G_{t_{i+1}}^{-},H_{t_{i+1}}^{-}\right)$. 

\section{\label{Sec4}Pricing method}

The fair value of the GMWB contract at time $t=0$ according to the
PH's subjective perspective, denoted by $\mathcal{V}\left(0,r_{0},P,P,P\right)$,
can be computed by moving backward in time. The terminal condition
is expressed by (\ref{eq:Terminal}). In order to proceed backward,
we have to solve the nonlinear implicit equation (\ref{eq:NIE}) in
$\left]t_{i},t_{i+1}\right[$ for $t_{i}=T-1,\dots,0$, and apply
relations (\ref{eq:V_t_minus}) and (\ref{eq:argmax}) to handle the
jumps due to withdrawals at each contract anniversary. 

With a similar approach, the initial fair value of the contract according
to the insurer's perspective, denoted by $\mathcal{U}\left(0,r_{0},P,P,P\right)$,
can be computed by starting from the terminal condition $\eqref{eq:Terminal-U}$,
by solving backward equation (\ref{eq:NIE-1}) and by applying relation
(\ref{eq:4.2-IN}). We observe that computing $\mathcal{U}\left(0,r_{0},P,P,P\right)$
requires the knowledge of the optimal withdrawals, which can be achieved
through the parallel computation of $\mathcal{V}\left(0,r_{0},P,P,P\right)$.

We stress out that the evaluation problems of $\mathcal{V}$ and $\mathcal{U}$
are four dimensional problems (plus the time variable) and this represents
a non-trivial challenge which requires an efficient numerical method
to be solved. 

\subsection{Problem discretisation \label{51}}

The variables that determine the state of the policy at any time are
the $G,$ $H,$ $X$ and $r$. To tackle the problem numerically,
we prefer to replace $r$ with $Y$, since the dynamics of $Y$ is
simpler and one can easily compute $r$ from $Y$ through (\ref{eq:rY}).
We consider a set of discrete values $\mathcal{\mathcal{G}}_{Y}$
for $Y$, $\mathcal{\mathcal{G}}_{X}$ for $X$, $\mathcal{\mathcal{G}}_{G}$
for $G$ and $\mathcal{\mathcal{G}}_{H}$ for $H$, and we define
a 4 dimensional grid $\mathcal{G}=\mathcal{\mathcal{G}}_{Y}\times\mathcal{\mathcal{G}}_{X}\times\mathcal{\mathcal{G}}_{G}\times\mathcal{\mathcal{G}}_{H}$. 

First of all, since the benefit base $G$ and the tax base $H$ are
non-negative values that do not exceed $P$, it is worth exploiting
an uniform partition of the interval $\left[0,P\right]$ to define
$\mathcal{\mathcal{G}}_{G}$ and $\mathcal{\mathcal{G}}_{H}$. In
particular, we set 
\begin{equation}
\mathcal{\mathcal{G}}_{G}=\left\{ g_{j}=\frac{j}{N_{G}}P,j=0,\dots,N_{G}\right\} 
\end{equation}
 and 
\begin{equation}
\mathcal{\mathcal{G}}_{H}=\left\{ h_{j}=\frac{j}{N_{H}}P,j=0,\dots,N_{H}\right\} ,
\end{equation}
where $N_{G}$ and $N_{H}$ are two positive integers.

As opposed to $G$ and $H$, the account value $X$ assumes non-negative
unbounded values. Anyway, because of withdrawals and fees applied
by the insurer, such a value should not grow too much during the life
of the policy. In fact, as observed by MacKay et al. \cite{mackay2017}
and by Bacinello and Zoccolan \cite{bacinello2019} in a similar context,
when the account value is very high there is a great incentive for
the PH to surrender the contract by withdrawing all the money. So,
following same principle of the spatial grid employed by Haentjens
and In't Hout \cite{haentjens2012}, we consider $\mathcal{\mathcal{G}}_{X}$
as a non-uniform distribution of points which is more dense where
the process $X$ is more likely to be. Specifically, we consider two
sets of points: the first set 
\begin{equation}
\mathcal{\mathcal{G}}_{X_{1}}=\left\{ x_{j}^{1}=2.5\cdot\frac{j}{N_{X_{1}}}P,j=0,\dots,N_{X_{1}}\right\} 
\end{equation}
 is made of $N_{X_{1}}+1$ uniformly distributed points between $0$
and $2.5\cdot P$ and the second one 
\begin{equation}
\mathcal{\mathcal{G}}_{X_{2}}=\left\{ x_{j}^{2}=2.5\cdot P\cdot\exp\left(\left(\ln\left(30\right)-\ln\left(2.5\right)\right)\frac{j}{N_{X_{2}}}\right),j=1,\dots,N_{X_{2}}\right\} 
\end{equation}
 is made of $N_{X_{2}}$ points which are distributed uniformly in
log between $2.5\cdot P$ and $30\cdot P$. Then, $\mathcal{\mathcal{G}}_{X}=\mathcal{\mathcal{G}}_{X_{1}}\cup\mathcal{\mathcal{G}}_{X_{2}}$
and we term $x_{j}$ the $j$-th point of $\mathcal{\mathcal{G}}_{X}$.
Moreover, for seek of simplicity, we consider $N_{X_{1}}=N_{X_{2}}$
and we term $N_{X}$ the number of elements of $\mathcal{\mathcal{G}}_{X}$.
We stress out that the coefficient $2.5$ and $30$ are determined
empirically in order to give accurate results and their small variations
do not produce impacts on the numerical results. 

Finally, the construction of the set $\mathcal{\mathcal{G}}_{Y}$
relies on the trinomial tree proposed by Goudenège et al. in \cite{goudenege2019}.
Such a tree defines a discrete Markov chain $\bar{Y}^{\Delta t}$
that matched the first two moments of the process $Y$. We set 
\begin{equation}
\mathcal{\mathcal{G}}_{Y}=\left\{ y_{j}=\frac{3}{2}\left(j-N_{Y}\right)\sigma_{Y}^{\Delta t},j=0,\dots2N_{Y}\right\} \label{eq:GY}
\end{equation}
 where $\sigma_{Y}^{\Delta t}$ is a positive coefficient that depends
on the standard deviation of the process $Y$ and $N_{Y}$ is a suitable
integer value, thus $\mathcal{\mathcal{G}}_{Y}$ is made of $2N_{Y}+1$
points uniformly distributed in $\left[-\frac{3}{2}N_{Y}\sigma_{Y}^{\Delta t};\frac{3}{2}N_{Y}\sigma_{Y}^{\Delta t}\right]$.
Appendix \ref{App0} presents technical details about the process
$\bar{Y}^{\Delta t}$, the coefficient $\sigma_{Y}^{\Delta t}$ and
the integer $N_{Y}$.

\subsection{Backward evaluation of $\mathcal{V}$}

Once the grid $\mathcal{G}$ has been build, we can start the computation
of the numerical approximation of $\mathcal{V}$ defined on $\mathcal{G}$
at any time $t_{i}$. In particular, for every policy anniversary
$t_{i}$, we compute a function $\mathcal{\bar{V}}_{i}^{+}:\mathcal{G}\rightarrow\mathbb{R}$
such that for any point $\left(y,x,g,h\right)$ of $\mathcal{G}$,
$\mathcal{\bar{V}}_{i}^{+}\left(y,x,g,h\right)$ approximates $\mathcal{V}^{+}\left(t_{i},y+\beta\left(t_{i}\right),x,g,h\right)$.
Moreover, we also compute a function $\mathcal{\bar{V}}_{i}^{-}:\mathcal{G}\rightarrow\mathbb{R}$
such that for any point $\left(y,x,g,h\right)$ of $\mathcal{G}$,
$\mathcal{\bar{V}}_{i}^{-}\left(y,x,g,h\right)$ approximates $\mathcal{V}^{-}\left(t_{i},y+\beta\left(t_{i}\right),x,g,h\right)$.
According to (\ref{eq:Final_value}), the terminal condition at each
point $\left(y,x,g,h\right)$ of $\mathcal{G}$ is given by: 
\begin{multline}
\mathcal{\bar{V}}_{T}^{-}\left(y,x,g,h\right)=\max\left\{ x,w_{T}\left(g\right)\right\} -\tau\min\left\{ w_{T}\left(g\right),\left(x-h\right)_{+}\right\} \\
-\tau\left(\left(w_{T}\left(g\right)-\left(x-h\right)_{+}\right)_{+}+\left(x-w_{T}\left(g\right)\right)_{+}-h\right)_{+},\label{eq:51}
\end{multline}
where $w_{T}\left(g\right)=\min\left\{ g^{W},g\right\} $. 

Suppose now the function $\mathcal{\bar{V}}_{i+1}^{-}$ to be known
on $\mathcal{G}$. Let us fix $\left(y,x,g,h\right)\in\mathcal{G}$
and let us focus on the computation of $\mathcal{\bar{V}}_{i}^{+}\left(y,x,g,h\right)$
by solving equation (\ref{eq:NIE}). In particular, following the
same approach employed by Moenig and Bauer \cite{moenig2016} under
the Black-Scholes model, one can verify that the solution of equation
(\ref{eq:NIE}) exists and is unique. Furthermore, according to equation
(\ref{eq:NIE}), $\mathcal{\bar{V}}_{i}^{+}\left(y,x,g,h\right)$
can be interpreted as the solution of a fixed point problem:
\begin{equation}
v=f\left(v\right),
\end{equation}
with 
\begin{equation}
f\left(v\right)=\mathbb{E}^{\mathbb{Q}}\left[e^{-\int_{t_{i}}^{t_{i+1}}Y_{s}+\beta\left(s\right)ds}\left(F+\frac{\kappa}{1-\kappa}\cdot\left(F-v\right)_{+}\right)\left|Y_{t_{i}}=y,X_{t_{i}}=x\right.\right],\label{eq:function_f}
\end{equation}
where $F$ stands for
\begin{equation}
F=q_{x+t_{i}}\left(X_{t_{i}+1}-\tau\left(X_{t_{i+1}}-h\right)_{+}\right)+p_{x+t_{i+1}}\mathcal{\bar{V}}_{i+1}^{-}\left(Y_{t_{i+1}},X_{t_{i+1}},g,h\right).
\end{equation}
Such a problem can be faced by fixed point iterations. The key point
consists in calculating the expected values that appears in (\ref{eq:function_f}).
In order to tackle such a problem, we employ a tree approach. Technical
details are explained in Appendix \ref{App1}.

Once the function $\mathcal{\bar{V}}_{i}^{+}$ is known, we can compute
$\mathcal{\bar{V}}_{i}^{-}$ by solving the optimal withdrawal problem
related to equation (\ref{eq:argmax}). So, let us fix again $\left(y,x,g,h\right)\in\mathcal{G}$
and let us focus on solving the following problem:
\begin{equation}
\mathcal{\bar{V}}_{i}^{-}\left(y,x,g,h\right)=\underset{w\in\left[0,W_{max}\right]}{\mbox{max}}\ \hat{f}\left(w\right)\label{eq:max}
\end{equation}
with 
\begin{equation}
\hat{f}\left(w\right)=\mathcal{\bar{V}}_{i}^{+}\left(y,x^{+}\left(w\right),g^{+}\left(w\right),h^{+}\left(w\right)\right)+\left(w-fee_{i}\left(w\right)-pen_{i}\left(w\right)-tax_{i}\left(w\right)\right),
\end{equation}
\begin{equation}
W_{max}=\max\left\{ x,\min\left\{ g^{W},g\right\} \right\} ,
\end{equation}
\begin{equation}
x^{+}\left(w\right)=\left(x-w\right)_{+}\label{57}
\end{equation}
\begin{equation}
g^{+}\left(w\right)=\begin{cases}
\left(g-w\right)_{+} & \text{, if }w\leq g^{W}\\
\left(\min\left\{ g-w,g\cdot\frac{x^{+}\left(w\right)}{x}\right\} \right)_{+} & \text{, if }w>g^{W}
\end{cases}
\end{equation}
\begin{equation}
h^{+}\left(w\right)=h-\left(w-\left(x-h\right)_{+}\right)_{+}
\end{equation}
\begin{equation}
fee_{i}\left(w\right)=s_{i}\cdot\left(w-\min\left\{ g^{W},g\right\} \right)_{+},\label{eq:fee-1}
\end{equation}
\begin{equation}
pen_{i}\left(w\right)=s^{g}\cdot\left(w-fee_{i}\left(w\right)\right)\cdot1_{\left\{ x+t_{i}<59.5\right\} },\label{eq:511}
\end{equation}
\begin{equation}
tax_{i}\left(w\right)=\tau\cdot\min\left\{ w-fee_{i}\left(w\right)-pen_{i}\left(w\right),\left(x-h\right)_{+}\right\} .\label{eq:tax-1}
\end{equation}
The resolution of (\ref{eq:max}) is not trivial as the function $\hat{f}$
has no smoothness properties. In particular $\hat{f}$ has singular
points, due to the presence of the positive part function, as well
as a discontinuity point, due to the function $g^{+}\left(w\right)$
at $w=g^{W}$. Therefore, we approach the solution of the maximization
problem (\ref{eq:max}) through a very simple approach: we evaluate
the target function in a set $W$ of points and record the maximum
value achieved on these points. In particular we consider $W$ as
the union of two sets, $W_{1}$ and $W_{2}$. The first set, $W_{1}=\left\{ n\cdot\Delta w,n\in\mathbb{N}\right\} \cap\left[0,W_{max}\right]$
is a set of uniformly distributed values, while the second set $W_{2}=\left\{ g^{W},g^{W}+10^{-6},W_{max}\right\} \cap\left[0,W_{max}\right]$
is a set of critical values that might not be included in $W_{1}$.
In particular, $g^{W}$ and $g^{W}+10^{-6}$ are considered in order
to handle the discontinuity at $g^{W}$. 

Evaluating the function $\hat{f}$ requires the calculation of the
function $\mathcal{\bar{V}}_{i}^{+}$ at $\left(y,x^{+}\left(w\right),g^{+}\left(w\right),h^{+}\left(w\right)\right)$
for any $w\in W$. These points may not belong the grid $\mathcal{G}$
as the values $x^{+}\left(w\right),g^{+}\left(w\right),h^{+}\left(w\right)$
may not belong to $\mathcal{G}_{X},\mathcal{G}_{G}$ and $\mathcal{G}_{H}$
respectively. So, in order to compute $\mathcal{\bar{V}}_{i}^{+}\left(y,x^{+}\left(w\right),g^{+}\left(w\right),h^{+}\left(w\right)\right)$,
interpolation on $\mathcal{G}$ of $\mathcal{\bar{V}}_{i}^{+}$ is
required. To this aim, we employ trilinear interpolation (Gomes et
al. \cite{gomes2009}).

\subsection{Backward evaluation of $\mathcal{U}$}

Once an approximation $\bar{\mathcal{V}}$ of $\mathcal{V}$ is available,
we can tackle the policy evaluation according to the insurer's perspective,
that is computing $\mathcal{U}$. Approximating $\mathcal{U}$ is
easier than approximating $\mathcal{V}$ because of two reasons. First
of all, the implicit nonlinear equation (\ref{eq:NIE}) is replaced
by an explicit equation (\ref{eq:NIE-1}). Secondly, the problem of
computing the best withdrawal has been already solved while approximating
$\mathcal{V}$, so we have just to recover the optimal withdrawals
already computed. 

Similarly to what we have done for $\mathcal{V}$, we consider a function
$\mathcal{\bar{U}}_{i}^{+}:\mathcal{G}\rightarrow\mathbb{R}$ such
that for any point $\left(y,x,g,h\right)$ of $\mathcal{G}$, $\mathcal{\bar{U}}_{i}^{+}\left(y,x,g,h\right)$
approximates $\mathcal{U}^{+}\left(t_{i},y+\beta\left(t_{i}\right),x,g,h\right)$
and a function $\mathcal{\bar{U}}_{i}^{-}:\mathcal{G}\rightarrow\mathbb{R}$
such that for any point $\left(y,x,g,h\right)$ of $\mathcal{G}$,
$\mathcal{\bar{U}}_{i}^{-}\left(y,x,g,h\right)$ approximates $\mathcal{U}^{-}\left(t_{i},y+\beta\left(t_{i}\right),x,g,h\right)$.
According to (\ref{eq:Final_U}), for any point $\left(y,x,g,h\right)$
of $\mathcal{G}$, the terminal condition is given by: 
\begin{equation}
\mathcal{\bar{U}}_{T}^{-}\left(y,x,g,h\right)=\max\left\{ x,\min\left\{ g^{W},g\right\} \right\} .\label{eq:513}
\end{equation}
 Suppose now the function $\mathcal{\bar{U}}_{i+1}^{-}$ to be known
on $\mathcal{G}$. Let us fix $\left(y,x,g,h\right)\in\mathcal{G}$
and let us focus on the computation of $\mathcal{\bar{U}}_{i}^{+}\left(y,x,g,h\right)$
by computing the following expression:
\begin{equation}
\mathcal{\bar{U}}_{i}^{+}\left(y,x,g,h\right)=\mathbb{E}^{\mathbb{Q}}\left[e^{-\int_{t_{i}}^{t_{i+1}}Y_{s}+\beta\left(s\right)ds}\left(q_{x+t_{i}}X_{t_{i+1}}+p_{x+t_{i+1}}\mathcal{\bar{U}}_{i+1}^{-}\left(Y_{t_{i+1}},X_{t_{i+1}},g,h\right)\right)\left|Y_{t_{i}}=y,X_{t_{i}}=x\right.\right].\label{eq:514}
\end{equation}
We compute such an expression by using the same tree approach employed
to compute (\ref{eq:function_f}). Please observe that in this case,
no fix point iterations are required because (\ref{eq:514}) gives
$\mathcal{U}_{i}^{+}$ through an explicit equation. 

Suppose now the function $\mathcal{\bar{U}}_{i}^{+}$ to be known
on $\mathcal{G}$. Let us fix $\left(y,x,g,h\right)\in\mathcal{G}$
and let us focus on the computation of $\mathcal{\bar{U}}_{i}^{-}\left(y,x,g,h\right)$.
Let $w_{i}$ be the maximum point for the problem (\ref{eq:max})
for the point $\left(y,x,g,h\right)\in\mathcal{G}$. The following
relation holds:
\begin{equation}
\mathcal{\bar{U}}_{i}^{-}\left(y,x,g,h\right)=\mathcal{\bar{U}}_{i}^{+}\left(y,x^{+}\left(w_{i}\right),g^{+}\left(w_{i}\right),h^{+}\left(w_{i}\right)\right)+\left(w-fee_{i}\left(w_{i}\right)-pen_{i}\left(w_{i}\right)\right)\label{515}
\end{equation}
where $x^{+}$, $g^{+}$, $h^{+}$, $fee_{i}$ and $pen_{i}$ are
defined as in (\ref{57})-(\ref{eq:511}). Also in this case, interpolation
is required and we employ again trilinear interpolation.

\subsection{Sketch of the algorithm}

We present the sketch of the algorithm to approximate the initial
fair contract values $\mathcal{V}\left(0,r_{0},P,P,P\right)$ and
$\mathcal{U}\left(0,r_{0},P,P,P\right)$.
\begin{enumerate}
\item Set the terminal values $\mathcal{\bar{V}}_{T}^{-}\left(y,x,g,h\right)$
and $\mathcal{\bar{U}}_{T}^{-}\left(y,x,g,h\right)$ according to
equations (\ref{eq:51}) and (\ref{eq:513}) for every point $\left(y,x,g,h\right)$
in $\mathcal{G}$.
\item For all $i=T-1,\dots1$
\begin{enumerate}
\item Compute $\mathcal{\bar{V}}_{i}^{+}\left(y,x,g,h\right)$ and $\mathcal{\bar{U}}_{i}^{+}\left(y,x,g,h\right)$
by solving equations (\ref{eq:function_f}) and in (\ref{eq:514})
for every point $\left(y,x,g,h\right)$ in $\mathcal{G}$;
\item Compute $\mathcal{\bar{V}}_{i}^{-}\left(y,x,g,h\right)$ by solving
equation (\ref{eq:max}) for every point $\left(y,x,g,h\right)$ in
$\mathcal{G}$;
\item Compute $\mathcal{\bar{U}}_{i}^{-}\left(y,x,g,h\right)$ by solving
equation (\ref{515}) for every point $\left(y,x,g,h\right)$ in $\mathcal{G}$;
\end{enumerate}
\item Compute $\mathcal{\bar{V}}_{0}^{+}\left(0,P,P,P\right)$ and $\mathcal{\bar{U}}_{0}^{+}\left(0,P,P,P\right)$
by solving equations (\ref{eq:function_f}) and (\ref{eq:514}).
\end{enumerate}
Values $\mathcal{\bar{V}}_{0}^{+}\left(0,P,P,P\right)$ and $\mathcal{\bar{U}}_{0}^{+}\left(0,P,P,P\right)$
approximate $\mathcal{V}\left(0,r_{0},P,P,P\right)$ and $\mathcal{U}\left(0,r_{0},P,P,P\right)$
respectively. We point out that the algorithm is fully parallelizable:
in fact the computations for every point in $\mathcal{G}$ are independent
of each other.

A common practice in VAs context (see for example Forsyth and Vetzal
\cite{forsyth2014}) consists in computing the fair policy cost $\varphi_{IN}^{*}$,
that is the particular value of $\varphi$ that makes the insurer's
initial value of the policy $\mathcal{U}\left(0,r_{0},P,P,P\right)$
equal to the net premium $P$. To this aim, the algorithm can be plugged
into the secant method to solve the equation $\mathcal{U}\left(0,r_{0},P,P,P\right)\left(\varphi\right)=P$.

\FloatBarrier

\section{Numerical Results \label{Sec6}}

In this Section we report the results of some numerical tests. Tables
\ref{tab:parameters_contract}, \ref{tab:parameters_model}, \ref{tab:parameters_numerical_methods}
report the parameters used in the analysis. Moreover, in order to
estimate the mortality and survival probabilities $q$ and $p$, we
employ the 2007 Period Life Table for the Social Security Area Population
for the USA \cite{Table}. Finally, we underline that these parameters,
with the exception of those for the interest rate process, are the
same employed by Moenig and Bauer \cite{moenig2016}.

\begin{table}
\begin{centering}
\begin{tabular}{lll}
\toprule 
Description & Parameter & \multicolumn{1}{l}{Value}\tabularnewline
\midrule
Age at inception & $x$ & $55$\tabularnewline
Premium & $P$ & $100$\tabularnewline
Years to maturity & $T$ & $15$\tabularnewline
Annual guaranteed amount  & $g^{W}$ & $7$\tabularnewline
Excess withdrawal fee & $s_{i}$ & $8\%,7\%,\dots,1\%,0\%,0\%,\dots$\tabularnewline
Fee rate & $\varphi$ & to be determined\tabularnewline
Income tax rate & $\tau$ & $0\%,$ or $30\%$\tabularnewline
Capital gain tax rate & $\kappa$ & $0\%,$ or $23\%$\tabularnewline
Early withdrawal penalty & $s^{g}$ & $10\%$\tabularnewline
\bottomrule
\end{tabular}
\par\end{centering}
\caption{\label{tab:parameters_contract}Parameter choices for the PH and contract
specifications.}
\end{table}

\begin{table}
\begin{centering}
\begin{tabular}{lll}
\toprule 
Description & Parameter & \multicolumn{1}{l}{Value}\tabularnewline
\midrule
Initial fund value & $S_{0}$ & $100$\tabularnewline
Fund volatility & $\sigma$ & $0.1$, $0.3$\tabularnewline
Initial interest rate  & $r_{0}$ & $0.03$, $0.05$\tabularnewline
Interest rate mean reversion speed  & $k$ & $1$\tabularnewline
Interest rate mean  & $\theta_{t}$ & $flat$\tabularnewline
Interest rate volatility & $\omega$ & $0.05,\ 0.1$\tabularnewline
Correlation  & $\rho$ & $0.2$\tabularnewline
\bottomrule
\end{tabular}
\par\end{centering}
\caption{\label{tab:parameters_model}Parameter choices for the Black-Scholes
Hull-White model.}
\end{table}

\begin{table}
\begin{centering}
\begin{tabular}{lll}
\toprule 
Description & Parameter & \multicolumn{1}{l}{Value}\tabularnewline
\midrule
Time step per year  & $N_{T}$ & $50$\tabularnewline
Points in $\mathcal{G}_{X}$  & $N_{X}$ & $500$\tabularnewline
Points in $\mathcal{G}_{G}$  & $N_{G}$ & $100$\tabularnewline
Points in $\mathcal{G}_{H}$  & $N_{H}$ & $100$\tabularnewline
Withdrawal step & $\Delta w$ & $1$\tabularnewline
\bottomrule
\end{tabular}
\par\end{centering}
\caption{\label{tab:parameters_numerical_methods}Parameter choices for the
numerical methods.}
\end{table}

\subsection{Computing the fair fee rate}

We start by computing the fair fee rate $\varphi_{IN}^{*}$ according
to the insurer's subjective valuation. In particular, we consider
some test cases with different values of the initial interest rate
$r_{0}$, the volatility of the interest rate $\omega$ and the volatility
of the underlying fund $\sigma$. Numerical results are reported in
Table \ref{tab:Results_1}. 

By comparing the two rows of Table \ref{tab:Results_1}, we observe
that, in all the considered cases, including taxation reduces the
fees that reduce the account value, that is the policy cost. We can
speculate that, the reason for this lies in the withdrawal strategy:
if taxation is applied, the optimal withdrawal strategy from PH's
perspective  (i.e. the strategy that maximizes value according to
his subjective view) does not coincide with the worst strategy according
to insurer's perspective (i.e. the strategy that maximizes his liability).
Moreover, we observe that the higher the interest rate volatility,
the greater the policy cost. This is probably due to the fact that,
by increasing the volatility of the interest rate, it is easier to
observe very low (or negative) interest rates which make replicating
the policy very expensive. Thus, both taxation and interest rate modeling
have a sensitive impact on policy evaluation.

\begin{table}
\begin{centering}
\begin{tabular}{cccccc}
\toprule 
 & \multicolumn{2}{c}{$r_{0}=0.03,\ \sigma=0.16$} &  & \multicolumn{2}{c}{$r_{0}=0.05,\ \sigma=0.19$}\tabularnewline
\cmidrule{2-3} \cmidrule{5-6} 
 & $\omega=0.05$ & $\omega=0.1$  &  & $\omega=0.05$ & $\omega=0.1$\tabularnewline
\midrule
\multirow{1}{*}{No Taxation} & {\small{}$69.35$} & {\small{}$94.81$} &  & {\small{}$41.91$} & {\small{}$56.97$}\tabularnewline
\multirow{1}{*}{With Taxation} & {\small{}$43.18$} & {\small{}$60.31$} &  & {\small{}$23.96$} & {\small{}$33.58$}\tabularnewline
\bottomrule
\end{tabular}
\par\end{centering}
\caption{\label{tab:Results_1}Fair fee rate $\varphi_{IN}^{*}$ (in basis
points) according to the insurer's subjective valuation, changing
the values of $r_{0},$ $\sigma$ and $\omega$.}
\end{table}

\subsection{Comparing policy initial values}

We compute now the PH's initial subjective policy value $\mathcal{V}\left(0,r_{0},P,P,P\right)$
with $\varphi$ equal to $\varphi_{IN}^{*}$, that is the break-even
fee, as in Table \ref{tab:Results_1}: this is the amount of money
the PH needs to replicate the policy on its own. Numerical results
are reported in Table \ref{tab:Results_2}. We observe that if no
taxation is applied, the subjective valuation of the PH equates the
subjective valuation of the insurer and the contract is fair for both
the two agents. Instead, when taxation is applied, the contract value
according to PH's subjective valuation increases because of the tax
regime applied to the policy, which is advantageous compared to the
tax regime applied to investment outside the policy, and thus in the
replicating portfolio. 

Now let us consider $\chi=3\%$ as the premium tax rate. With such
a premium tax rate, the gross premium $GP$ that the insurer requires
to cover all the costs is $103.09$, so that the net premium $P$
is $100$ and the contract is fair for the insurer. According to Table
\ref{tab:Results_2}, in all the considered cases, the customer will
be willing to pay much more than $103.90$ to buy the policy: for
example, if $r_{0}=0.03$, $\sigma=0.16$ and $\omega=0.05$ then
the PH's subjective valuation of the policy is $110.13$. Therefore,
if the insurer sets a sale price between $103.09$ and $110.13$,
then the sale will be advantageous for both the PH and the insurer.

The model considered allows us to recreate a framework that makes
VAs particularly interesting to customers: although taxes are applied
on the earnings of the policy, the tax regime is particularly favorable
for this type of product. Therefore the GMWB policy is attractive
for the customer and profitable for the insurer.

\begin{table}
\begin{centering}
\begin{tabular}{cccccc}
\toprule 
 & \multicolumn{2}{c}{$r_{0}=0.03,\ \sigma=0.16$} &  & \multicolumn{2}{c}{$r_{0}=0.05,\ \sigma=0.19$}\tabularnewline
\cmidrule{2-3} \cmidrule{5-6} 
 & $\omega=0.05$ & $\omega=0.1$ &  & $\omega=0.05$ & $\omega=0.1$\tabularnewline
\midrule
\multirow{1}{*}{No Taxation} & {\small{}$100.00$} & {\small{}$100.00$} &  & {\small{}$100.00$} & {\small{}$100.00$}\tabularnewline
\multirow{1}{*}{With Taxation} & {\small{}$110.13$} & {\small{}$110.69$} &  & {\small{}$114.97$} & {\small{}$115.91$}\tabularnewline
\bottomrule
\end{tabular}
\par\end{centering}
\caption{\label{tab:Results_2}Fair initial option value $\mathcal{V}\left(0,r_{0},P,P,P\right)$
according to PH's subjective valuation, while considering $\varphi=\varphi_{IN}^{*}$
as in Table \ref{tab:Results_1}, changing the values of $r_{0},$
$\sigma$ and $\omega$.}
\end{table}

\FloatBarrier

\subsection{Comparing the PH withdrawals with and without taxation}

\begin{figure}
\begin{centering}
\includegraphics[width=1\textwidth,trim={2.5cm 0cm 2.5cm 0cm},clip]{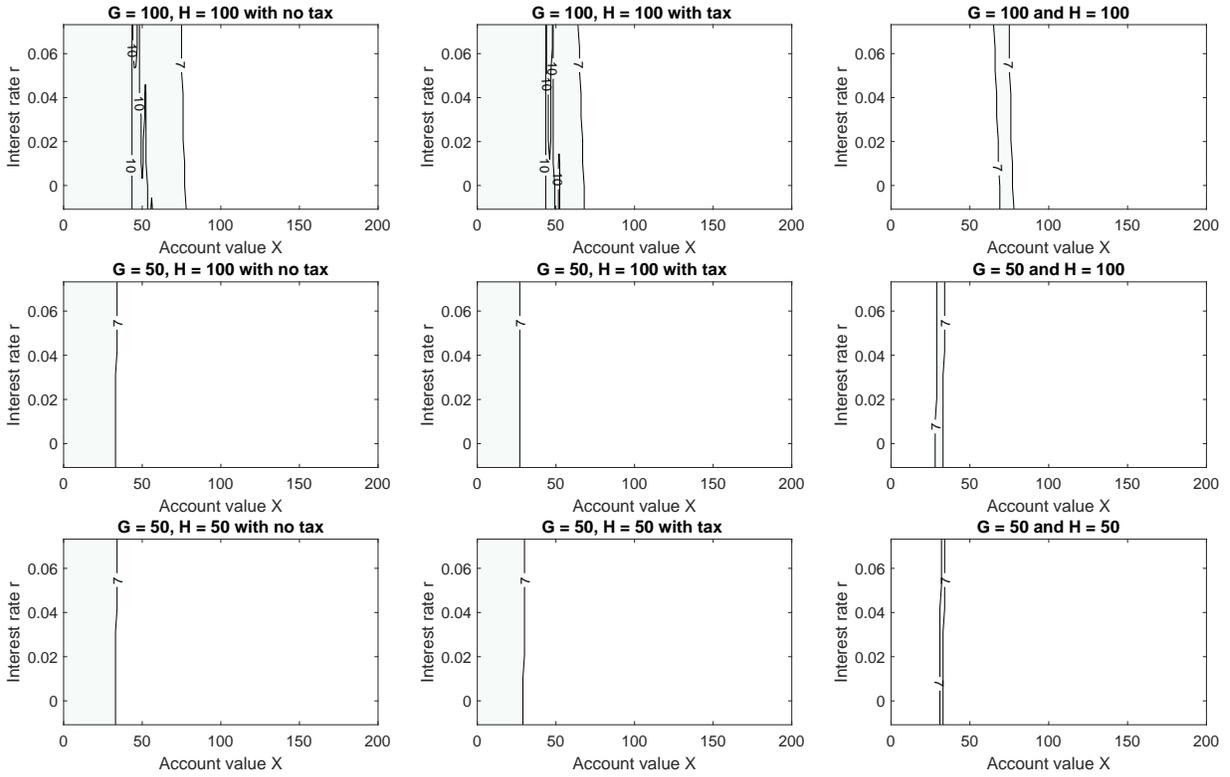}
\par\end{centering}
\caption{\label{F4}In the first two columns, the contour plot with respect
to $X_{4}^{-}$ (x-axis) and $r_{4}$ (y-axis) of the optimal withdrawal
$w_{4}$ at time $t_{4}=4$ without or applying taxation.  In the
third column the difference between the optimal withdrawals without
and with taxation reported in the first two columns. Green areas denote
positive values.}
\end{figure}

\begin{figure}
\begin{centering}
\includegraphics[width=1\textwidth,trim={2.5cm 0cm 2.5cm 0cm},clip]{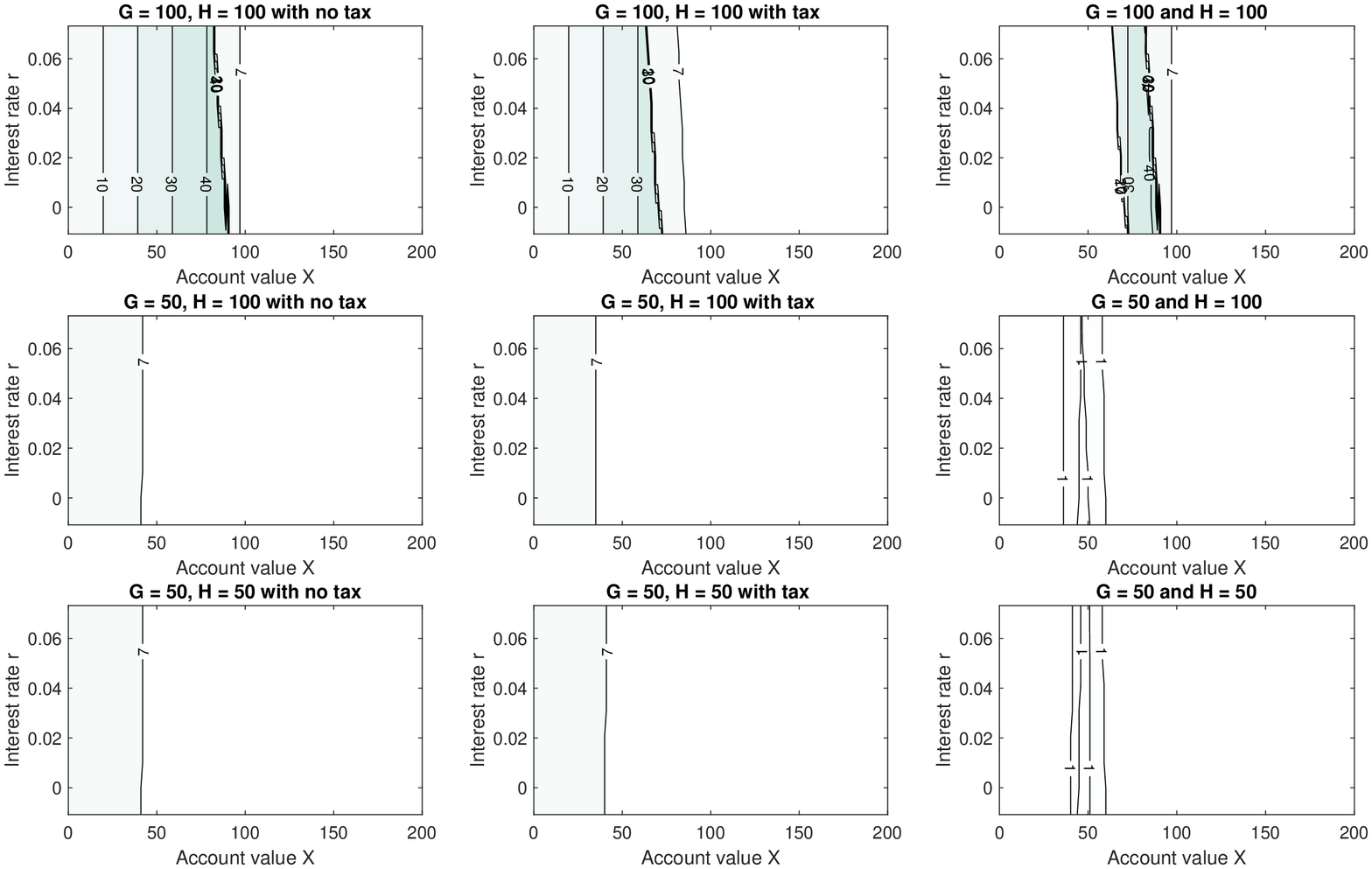}
\par\end{centering}
\caption{\label{F8}Same contour plots as in Figure \ref{F4}, but considering
$t_{8}$ in place of $t_{4}$.}
\end{figure}

\begin{figure}
\begin{centering}
\includegraphics[width=1\textwidth,trim={2.5cm 0cm 2.5cm 0cm},clip]{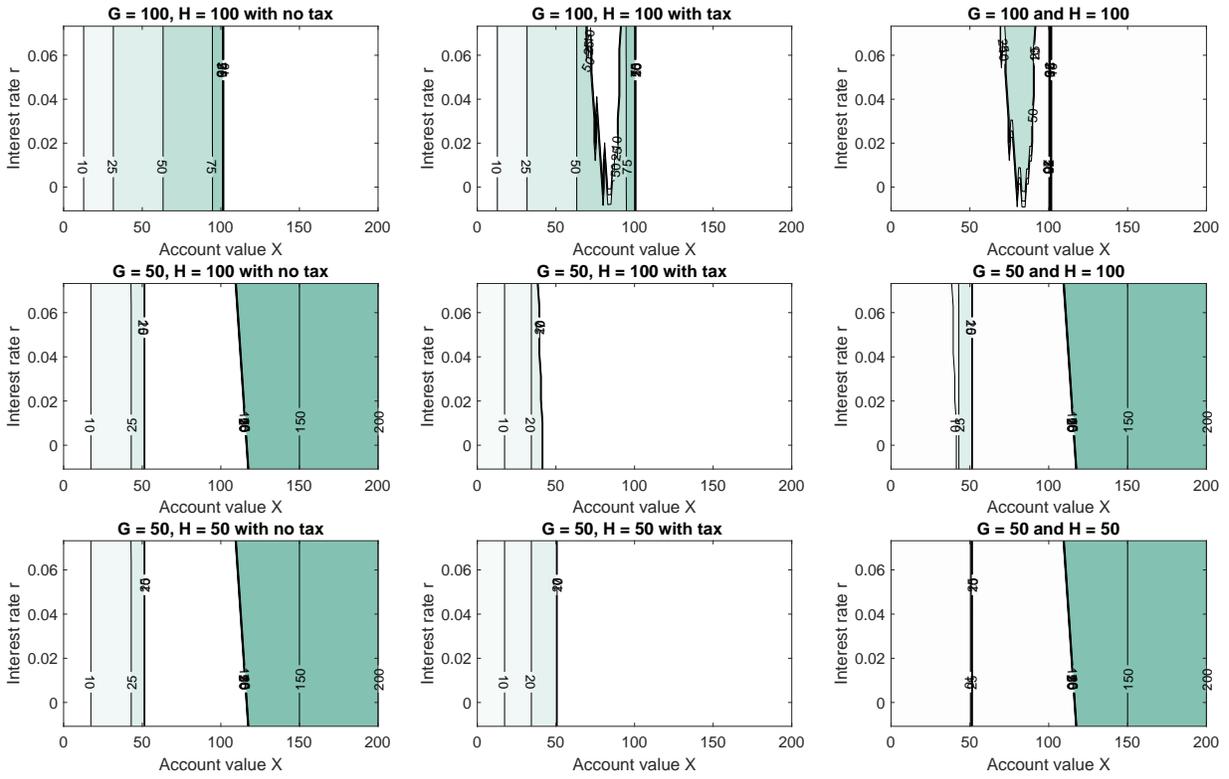}
\par\end{centering}
\caption{\label{F12}Same contour plots as in Figure \ref{F4}, but considering
$t_{12}$ in place of $t_{4}$.}
\end{figure}

The last numerical test we propose consists in comparing the optimal
withdrawals performed by the PH while considering or not taxation.
In particular, we consider the case with $r_{0}=0.03$, $\sigma=0.16$
and $\omega=0.05$ (results for other parameters combinations are
similar). Moreover, the value of $\varphi$ is set as the break-even
fee with taxes, that is $\varphi_{IN}^{*}=43.18$ basis points. Optimal
withdrawal amounts $w_{i}$ at different anniversaries are reported
in Figures \ref{F4}, \ref{F8} and \ref{F12}. The first column represents
the optimal amount as a function of $X_{t_{i}}^{-}$ and $r_{t_{i}}$,
by considering different values for $G$ and $H$ and a zero tax rate,
while in the second column by considering both a positive income tax
rate and a positive capital gain tax rate. The area where the color
is darker identifies higher withdrawals. Finally, in the third column,
the difference between the optimal amount without taxation and with
taxation. Here, green areas indicate that withdrawals without tax
are higher, while red areas (not visible) indicate that withdrawals
with tax are higher.

We can observe that the optimal amount depends on all the considered
parameters. In particular, the withdrawal strategy may significantly
change according to the actual interest rate, as shown in Figure \ref{F12}
for $G_{t_{12}}=H_{t_{12}}=100$. As far as the impact of taxation
on withdrawal strategy is concerned, we find the same effect observed
by Moenig and Bauer \cite{moenig2016}: when taxation is applied,
the PH withdraws less than when taxation is not considered. In fact,
for all the numerical cases considered, the last columns in Figures
\ref{F4}, \ref{F8} and \ref{F12} show only positive values, which
means the amount withdrawn with no taxation is higher. As observed
in \cite{moenig2016}, in the absence of taxes, as the account value
increases more and more, the PH is motivated to withdraw money instead
of leaving it in the policy where it is reduced by fees. Conversely,
if taxation is applied, withdrawals are taxed as ordinary income and
they are subject to capital gain tax if invested in other products.
Therefore it is more convenient for the PH not to withdraw the money,
letting it grow within the policy. Such a difference in the withdrawal
strategy is particularly clear in Figure \ref{F12}: if $G=H=50$,
then the PH withdraws large amount of money when $X$ is high if taxation
is neglected, whereas no money if taxation is applied.

\FloatBarrier

\section{Conclusions}

In this paper we have investigate the impact of taxation on a GMWB
Variable Annuity when stochastic interest rate is considered. We modeled
taxation following the approach of Moenig and Bauer \cite{moenig2016}:
we have considered a subjective risk-neutral valuation methodology
that considers differences in the taxation for both different products
and market agents. Moreover, we have modeled stochastic interest rate
through the Hull-White model. This analysis combines the effects of
taxation and of the variable interest rate which, as already shown
separately in other research work, can have a significant impact on
the withdrawal choices and thus on the hedging costs. This analysis
has been possible thanks to the use of an efficient numerical method
based on a tree approach (Goudenège et al. \cite{goudenege2019}).
Numerical results show many interesting facts. First of all both,
taxation and interest rate modeling can have a relevant impact on
policy evaluation: the break-even fee can change of several basis
points when the parameters of these two factors change. Then, applying
different taxation to insurer and to policy holder can lead to different
policy evaluation: in particular policy holder's valuation is higher
than insurer's valuation and this makes buying and selling the policy
convenient for both the two agents. Moreover, numerical tests show
that taxation clearly impacts on withdrawal strategy: it discourages
the policy holder to perform withdrawals. This is useful to match
theoretical prices to those actually observed on the real market.
In conclusion, the model presented here represents an important extension
in the evaluation of GMWB type policies.

\FloatBarrier

\appendix

\section{Markov chain to approximate $Y$\label{App0}}

In this Appendix we explain how to design a discrete time Markov chain
that approximates the process $Y$, based on the trinomial tree introduced
in \cite{goudenege2019}. First of all, we consider a partition of
the time interval $\left[0,T\right]$ in $T\cdot N_{T}$ sub-intervals,
that is $N_{T}$ per year. We define $\Delta t=\frac{1}{N_{T}}$ as
the time increment and we term $\bar{t}_{n}=n\cdot\Delta t$ the $n$-th
time step for $n=0,\dots,T\cdot N_{T}$. Please observe that policy
anniversaries $t_{0},\dots,t_{T}$ are included in the time steps
$\bar{t}_{0},\bar{t}_{1}\dots,\bar{t}_{T\cdot N_{T}}$ and in particular
$t_{i}=\bar{t}_{i\cdot N_{T}}=i$. We consider the set $\mathcal{Y}$
given by
\begin{equation}
\mathcal{Y}=\left\{ \upsilon_{j}=\frac{3}{2}j\sigma_{Y}^{\Delta t},j\in\mathbb{Z}\right\} ,
\end{equation}
 where the coefficient $\sigma_{Y}^{\Delta t}$ is the standard deviation
of the random variable $Y_{\bar{t}_{n+1}}-Y_{\bar{t}_{n}}$ (which
is the same for all $\bar{t}_{n}$ values) and it is given by
\begin{equation}
\sigma_{Y}^{\Delta t}=\omega\sqrt{\frac{1-\exp\left(-2k\cdot\Delta t\right)}{2k}}.\label{eq:syt}
\end{equation}
We define now a discrete time Markov chain $\bar{Y}^{\Delta t}=\left\{ \bar{Y}_{n}^{\Delta t},n=0,\dots,T\cdot N_{T}\right\} $
whose state space is an opportune subset of $\mathcal{Y}$ and that
matches the first two moments of the process $Y=\left\{ Y_{t},0\leq t\leq T\right\} $.
The process $\bar{Y}^{\Delta t}$ is designed so it weakly converges
to the process $Y$: in particular $\bar{Y}_{n}^{\Delta t}$ converges
to $Y_{\bar{t}_{n}}$. 

The initial value is $\bar{Y}_{0}^{\Delta t}=\upsilon_{0}=0$, so
$\bar{Y}_{0}^{\Delta t}=Y_{0}$. Now, let us fix a value $n\in\left\{ 0,\dots,T\cdot N_{T}-1\right\} $
and suppose $\bar{Y}_{n}^{\Delta t}=\upsilon_{m}$ for a certain integer
$m\in\mathbb{Z}$. Let 
\begin{equation}
\mu_{Y}^{\Delta t}\left(\upsilon_{m}\right)=\mathbb{E}\left[Y_{\bar{t}_{n+1}}\left|Y_{\bar{t}_{n}}=\upsilon_{m}\right.\right]=\upsilon_{m}\cdot\exp\left(-k\cdot\Delta t\right)
\end{equation}
 be the expected value of the random variable $Y_{\bar{t}_{n+1}}\left|Y_{\bar{t}_{n}}=\upsilon_{m}\right.$.
We term
\begin{equation}
j_{A}=\mbox{ceil}\left[\frac{2}{3\sigma_{Y}^{\Delta t}}\mu_{Y}^{\Delta t}\left(\upsilon_{m}\right)\right]
\end{equation}
the index of the first element of $\mathcal{Y}$ whose value is bigger
than the expected value of the process $Y_{\bar{t}_{n+1}}\left|Y_{\bar{t}_{n}}=\upsilon_{m}\right.$.
Moreover, we also consider these three indices:
\begin{equation}
j_{B}=j_{A}-1,\ j_{C}=j_{A}+1,\ j_{D}=j_{A}-2.
\end{equation}
In particular, if we define the variables 
\begin{equation}
\Delta^{A}=\upsilon_{j_{A}}-\mu_{Y}^{\Delta t}\left(\upsilon_{m}\right)
\end{equation}
 and 
\begin{equation}
\Delta^{B}=\mu_{Y}^{\Delta t}\left(\upsilon_{m}\right)-\upsilon_{j_{B}}
\end{equation}
 then $0\le\Delta^{A}<\frac{3}{2}\sigma_{Y}^{\Delta t}$ and $0<\Delta^{B}\leq\frac{3}{2}\sigma_{Y}^{\Delta t}$.

There are two alternatives for the future states of the process $\bar{Y}^{\Delta t}$:
it can move from $\upsilon_{m}$ either to $\upsilon_{j_{A}}$, $\upsilon_{j_{B}}$,
$\upsilon_{j_{C}}$, or to $\upsilon_{j_{A}}$, $\upsilon_{j_{B}}$,
$\upsilon_{j_{D}}$. Transition probabilities $p_{A},p_{B},p_{C},p_{D}$
for both these two alternatives are stated in Table \ref{tab:Transition-probabilities-for}.
In particular, it is possible to prove that if $0\leq\Delta^{A}\leq\frac{\sqrt{5}}{2}\sigma_{Y}^{\Delta t}$
then $p_{A,}p_{B},p_{C}\in\left[0,1\right],$ while if $\frac{3-\sqrt{5}}{2}\sigma_{Y}^{\Delta t}\leq\Delta^{A}<\frac{3}{2}\sigma_{Y}^{\Delta t}$
then $p_{A,}p_{B},p_{D}\in\left[0,1\right]$. Since $\frac{3-\sqrt{5}}{2}\sigma_{Y}^{\Delta t}<\frac{\sqrt{5}}{2}\sigma_{Y}^{\Delta t}$,
at least one of the two sets of probabilities is well defined. Moreover,
transition probabilities in Table \ref{tab:Transition-probabilities-for}
have been computed in order to match the first two moments of the
process $Y$: this means that the random vectors $\bar{Y}_{n+1}^{\Delta t}-\bar{Y}_{n}^{\Delta t}$
and $Y_{\bar{t}_{n+1}}-Y_{\bar{t}_{n}}$, given $\bar{Y}_{n}^{\Delta t}=Y_{\bar{t}_{n}}=y_{m}$,
have the same mean and variance.

\begin{table}
\begin{centering}
\begin{tabular}{ccc}
 & Transition to $\upsilon_{j_{A}},\upsilon_{j_{B}},\upsilon_{j_{C}}$ & Transition to $\upsilon_{j_{A}},\upsilon_{j_{B}},\upsilon_{j_{D}}$\tabularnewline
\midrule
$p_{A}$ & $\frac{5\left(\sigma_{Y}^{\Delta t}\right)^{2}-4\left(\Delta^{A}\right)^{2}}{9\left(\sigma_{Y}^{\Delta t}\right)^{2}}$ & $\frac{2\left(\Delta^{B}\right)^{2}+3\left(\sigma_{Y}^{\Delta t}\right)\left(\Delta^{B}\right)+2\left(\sigma_{Y}^{\Delta t}\right)^{2}}{9\left(\sigma_{Y}^{\Delta t}\right)^{2}}$\tabularnewline
$p_{B}$ & $\frac{2\left(\Delta^{A}\right)^{2}+3\left(\sigma_{Y}^{\Delta t}\right)\left(\Delta^{A}\right)+2\left(\sigma_{Y}^{\Delta t}\right)^{2}}{9\left(\sigma_{Y}^{\Delta t}\right)^{2}}$ & $\frac{5\left(\sigma_{Y}^{\Delta t}\right)^{2}-4\left(\Delta^{B}\right)^{2}}{9\left(\sigma_{Y}^{\Delta t}\right)^{2}}$\tabularnewline
$p_{C}$ & $\frac{2\left(\Delta^{A}\right)^{2}-3\left(\sigma_{Y}^{\Delta t}\right)\left(\Delta^{A}\right)+2\left(\sigma_{Y}^{\Delta t}\right)^{2}}{9\left(\sigma_{Y}^{\Delta t}\right)^{2}}$ & $0$\tabularnewline
$p_{D}$ & $0$ & $\frac{2\left(\Delta^{B}\right)^{2}-3\left(\sigma_{Y}^{\Delta t}\right)\left(\Delta^{B}\right)+2\left(\sigma_{Y}^{\Delta t}\right)^{2}}{9\left(\sigma_{Y}^{\Delta t}\right)^{2}}$\tabularnewline
\bottomrule
\end{tabular}
\par\end{centering}
\caption{\label{tab:Transition-probabilities-for}Transition probabilities
for the process $\bar{Y}^{\Delta t}$.}
\end{table}

The choice between the two alternatives \-- $\upsilon_{j_{A}}$,
$\upsilon_{j_{B}}$, $\upsilon_{j_{C}}$ or $\upsilon_{j_{A}}$, $\upsilon_{j_{B}}$,
$\upsilon_{j_{D}}$ \-- is made in order to reduce the number of
points connected with $\upsilon_{0}$, which is the starting point.
Since $Y$ reverts to $0$, it is sufficient, when possible, to choose
the set with the points closest to $\upsilon_{0}$. Specifically,
if $\Delta^{A}<\frac{3-\sqrt{5}}{2}\sigma_{Y}^{\Delta t}$ then $\bar{Y}^{\Delta t}$
can only move to $\upsilon_{j_{A}}$, $\upsilon_{j_{B}}$, $\upsilon_{j_{C}}$
(in fact at least one among $p_{A,}p_{B}$ and $p_{D}$ is not in
$\left[0,1\right]$). If $\frac{\sqrt{5}}{2}\sigma_{Y}^{\Delta t}<\Delta^{A}$
then $\bar{Y}^{\Delta t}$ can only move to $\upsilon_{j_{A}}$, $\upsilon_{j_{B}}$,
$\upsilon_{j_{D}}$ (in fact at least one among $p_{A,}p_{B}$ and
$p_{C}$ is not in $\left[0,1\right]$). Finally, if $\frac{3-\sqrt{5}}{2}\sigma_{Y}^{\Delta t}\leq\Delta^{A}\leq\frac{\sqrt{5}}{2}\sigma_{Y}^{\Delta t}$
both choices are admissible: if $\left|\upsilon_{j_{C}}\right|\leq\left|\upsilon_{j_{D}}\right|$,
then $\bar{Y}^{\Delta t}$ moves to $\upsilon_{j_{A}},\upsilon_{j_{B}},\upsilon_{j_{C}}$
otherwise to $\upsilon_{j_{A}},\upsilon_{j_{B}},\upsilon_{j_{D}}$.

The state space of $\bar{Y}^{\Delta t}$ is the connected component
of $\upsilon_{0}$, that is the set $\mathcal{Y}_{0}\subset\mathcal{Y}$
of points that the process $\bar{Y}^{\Delta t}$ can reach. Taking
advantage of the symmetry and mean reversion properties of the process
$Y$, and thus of $\bar{Y}^{\Delta t},$ one can prove that $\mathcal{Y}_{0}=\left\{ \upsilon_{j},j=-N_{Y},\dots,N_{Y}\right\} $
where $N_{Y}$ is an integer. Moreover, by exploiting the definition
of $\bar{Y}^{\Delta t}$, one can prove that 
\begin{equation}
N_{Y}\leq\frac{\left(3-\sqrt{5}\right)e^{k\cdot\Delta t}}{3\left(e^{k\cdot\Delta t}-1\right)}+1,
\end{equation}
thus $N_{Y}\leq\frac{3-\sqrt{5}}{3k}\cdot N_{T}$ as $N_{T}\rightarrow+\infty$.

Finally, we stress out that $\bar{Y}^{\Delta t}$ matches the first
two moments of the process $Y$, thus weak convergence for $N_{T}\rightarrow+\infty$
is guaranteed ant it can be proved as done by Nelson and Ramaswamy
\cite{nelson1990}. 

To conclude, we observe that the set $\mathcal{\mathcal{G}}_{Y}$
defined in (\ref{eq:GY}) is equal to $\mathcal{Y}_{0}$: the only
difference concerns the indexing of the elements and in particular
$y_{j}=\upsilon_{j-N_{Y}}$ for $j\in\left\{ 0,\dots,2N_{Y}\right\} $. 

\section{Computing expected value (\ref{eq:function_f}) \label{App1}}

\subsection{The binomial tree approach}

In this Appendix we explain how to efficiently compute the expectation
in (\ref{eq:function_f}). Such a computation can be seen as a particular
case of a more general problem: computing
\begin{equation}
E=\mathbb{E}^{\mathbb{Q}}\left[e^{-\int_{t_{i}}^{t_{i+1}}Y_{s}+\beta\left(s\right)ds}\phi\left(Y_{t_{i+1}},X_{t_{i+1}},g,h\right)\left|Y_{t_{i}}=y,X_{t_{i}}=x\right.\right],\label{eq:target}
\end{equation}
where $\phi:A\subset\mathbb{R}^{4}\rightarrow\mathbb{R}$ is a given
function. Moreover $t_{i+1}$ and $t_{i}$ are two consecutive policy
anniversaries times and thus $t_{i+1}-t_{i}=1$. Furthermore, $y\in\mathcal{G}_{Y}$
and $x\in\mathcal{G}_{X}$.

First of all, let us consider the Gaussian vector $\varLambda=\left(\varLambda_{1},\varLambda_{2},\varLambda_{3}\right)^{\top}$given
by
\begin{equation}
\varLambda=\left(Y_{t_{i+1}},\ln\left(X_{t_{i+1}}\right),\int_{t_{i}}^{t_{i+1}}Y_{s}+\beta\left(s\right)ds\right)^{\top}\left|Y_{t_{i}}=y,X_{t_{i}}=x\right..\label{eq:82}
\end{equation}
It is possible to prove that the mean vector $\mu$ of $\varLambda$
is given by
\[
\mu=\left(\mu_{1,}\mu_{2},\mu_{3}\right)^{\top},
\]
where 
\begin{equation}
\mu_{1}=y\cdot e^{-k},\ \mu_{2}=\ln\left(x\right)+\mu_{3}-\varphi-\frac{1}{2}\sigma^{2},\ \mu_{3}=y\frac{1-e^{-k}}{k}+\int_{t_{i}}^{t_{i+1}}\beta\left(s\right)ds.
\end{equation}
 Moreover, the covariance matrix $\Pi$ of $\varLambda$ is given
by
\begin{equation}
\Pi=\left(\begin{array}{ccc}
\Pi_{11} & \Pi_{12} & \Pi_{13}\\
\Pi_{12} & \Pi_{22} & \Pi_{23}\\
\Pi_{13} & \Pi_{23} & \Pi_{33}
\end{array}\right),
\end{equation}
with
\begin{equation}
\Pi_{11}=\frac{1}{2}\omega^{2}\frac{(1-e^{-2k})}{k},
\end{equation}
\begin{equation}
\Pi_{33}=\left(\frac{\omega}{k}\right)^{2}\left(1+\frac{2e^{-k}}{k}-\frac{e^{-2k}}{2k}-\frac{3}{2k}\right),
\end{equation}
\begin{equation}
\Pi_{22}=\Pi_{33}+\sigma^{2}+2\sigma\rho\frac{\omega}{k}\left(1-\frac{1-e^{-k}}{k}\right),
\end{equation}
\begin{equation}
\Pi_{13}=\frac{\omega^{2}}{2}\left(\frac{1-e^{-k}}{k}\right)^{2},
\end{equation}
\begin{equation}
\Pi_{12}=\Pi_{13}+\sigma\rho\omega\frac{1-e^{-k}}{k},
\end{equation}
\begin{equation}
\Pi_{23}=\Pi_{33}+\sigma\rho\frac{\omega}{k}\left(1-\frac{1-e^{-k}}{k}\right).
\end{equation}
Let $\Gamma$ be the lower triangular Cholesky decomposition of $\Pi$,
and suppose $\tilde{G}=\left(\tilde{G}_{1},\tilde{G}_{2},\tilde{G}_{3}\right)^{\top}$
to be a Gaussian standard vector. So, the random vector $\mu+\Gamma\tilde{G}$
has the same law of $\Lambda$. Following the same approach of Ekvall
\cite{ekvall1996} for multidimensional simulation, we can develop
a binomial tree method to compute (\ref{eq:target}). Such a method
exploits 3 independent binomial approximations of $\tilde{G}_{1}$,
$\tilde{G}_{2}$ and $\tilde{G}_{3}$. In particular, we consider
the binomial random variable 
\begin{equation}
B^{N}\sim Bi\left(N_{T},\frac{1}{2}\right)
\end{equation}
and define 
\begin{equation}
\hat{G}^{N_{T}}=\frac{B^{N_{T}}-\frac{N_{T}}{2}}{\sqrt{\frac{N_{T}}{4}}}.
\end{equation}
It is well known that $\hat{G}^{N_{T}}$ converges in distribution
to a standard normal distribution so, if $\hat{G}_{1}^{N_{T}},\hat{G}_{2}^{N_{T}},\hat{G}_{3}^{N_{T}}$
are i.i.d. random variables that have the same law of $\hat{G}^{N_{T}}$,
then the vector $\hat{\Lambda}^{N_{T}}=\left(\hat{\Lambda}_{1}^{N_{T}},\hat{\Lambda}_{2}^{N_{T}},\hat{\Lambda}_{3}^{N_{T}}\right)^{\top}$given
by
\begin{equation}
\hat{\Lambda}^{N_{T}}=\mu+\Gamma\left(\hat{G}_{1}^{N_{T}},\hat{G}_{2}^{N_{T}},\hat{G}_{3}^{N_{t}}\right)^{\top}
\end{equation}
converges in distribution to $\varLambda$. Let $\left\{ \hat{g}^{0},\dots,\hat{g}^{N_{T}}\right\} $
be the support of $\hat{G}^{N_{T}}$ and let 
\begin{equation}
\hat{p}^{m}=\mathbb{P}\left[\hat{G}^{N_{t}}=\hat{g}^{l}\right]=\left(\begin{array}{c}
N_{T}\\
l
\end{array}\right)\left(\frac{1}{2}\right)^{N_{T}}
\end{equation}
 for $l=0,\dots,N_{T}$ be the associated probabilities. Let $\xi_{1},\xi_{2},\xi_{3}$
be three integers in $\left\{ 0,\dots,N_{T}\right\} $ and let $\left(\hat{\lambda}_{1}^{\xi_{1}},\hat{\lambda}_{2}^{\xi_{1},\xi_{2}},\hat{\lambda}_{3}^{\xi_{1},\xi_{2},\xi_{3}}\right)^{\top}$
be the vector defined by
\begin{equation}
\left(\hat{\lambda}_{1}^{\xi_{1}},\hat{\lambda}_{2}^{\xi_{1},\xi_{2}},\hat{\lambda}_{3}^{\xi_{1},\xi_{2},\xi_{3}}\right)^{\top}=\mu+\Gamma\left(\hat{g}^{\xi_{1}},\hat{g}^{\xi_{2}},\hat{g}^{\xi_{3}}\right)^{\top}.
\end{equation}
Please observe that, since $\Gamma$ is lower triangular, $\hat{\lambda}_{1}^{\xi_{1}}$
does not depend on $\hat{g}^{\xi_{1}}$ and $\hat{g}^{\xi_{2}}$,
while $\hat{\lambda}_{2}^{\xi_{1},\xi_{2}}$ does not depend on $\hat{g}^{\xi_{3}}$.
Moreover 
\begin{equation}
\mathbb{P}\left[\hat{\Lambda}_{1}^{N}=\hat{\lambda}_{1}^{\xi_{1}},\hat{\Lambda}_{2}^{N}=\hat{\lambda}_{2}^{\xi_{1},\xi_{2}},\hat{\Lambda}_{3}^{N}=\hat{\lambda}_{3}^{\xi_{1},\xi_{2},\xi_{3}}\right]=\hat{p}^{\xi_{1}}\cdot\hat{p}^{\xi_{2}}\cdot\hat{p}^{\xi_{3}}.
\end{equation}
In order to approximate (\ref{eq:target}), we replace $Y_{t_{i+1}},\ln\left(X_{t_{i+1}}\right)$
and $\int_{t_{i}}^{t_{i+1}}Y_{s}+\beta\left(s\right)ds$ with $\hat{\Lambda}_{1}^{N},\hat{\Lambda}_{2}^{N}$
and $\hat{\Lambda}_{3}^{N}$ respectively. We obtain
\begin{align}
\hat{E} & =\mathbb{E}\left[e^{-\hat{\Lambda}_{3}^{N_{T}}}\phi\left(\hat{\Lambda}_{1}^{N_{T}},\hat{\Lambda}_{2}^{N_{T}},g,h\right)\right]\label{eq:813}\\
 & =\sum_{\xi_{1}=0}^{N_{T}}\sum_{\xi_{2}=0}^{N_{T}}\sum_{\xi_{3}=0}^{N_{T}}\hat{p}^{\xi_{1}}\hat{p}^{\xi_{2}}\hat{p}^{\xi_{3}}\exp\left(-\hat{\lambda}_{3}^{\xi_{1},\xi_{2},\xi_{3}}\right)\phi\left(\hat{\lambda}_{1}^{\xi_{1}},\exp\left(\hat{\lambda}_{2}^{\xi_{1},\xi_{2}}\right),g,h\right)\\
 & =\sum_{\xi_{1}=0}^{N_{T}}\hat{p}^{\xi_{1}}\sum_{\xi_{2}=0}^{N_{T}}\hat{p}^{\xi_{2}}\phi\left(\hat{\lambda}_{1}^{\xi_{1}},\exp\left(\hat{\lambda}_{2}^{\xi_{1},\xi_{2}}\right),g,h\right)\sum_{\xi_{3}=0}^{N_{T}}\hat{p}^{\xi_{3}}\exp\left(-\hat{\lambda}_{3}^{\xi_{1},\xi_{2},\xi_{3}}\right)\label{eq:87}
\end{align}
Such an expression converges to $E$ thanks to the properties of convergence
in distribution for expected values (see for example Pollard \cite{pollard2012convergence}).
Please observe that, by leaving the random variable $\int_{t_{i}}^{t_{i+1}}Y_{s}+\beta\left(s\right)ds$
as the third component in $\Lambda$, the two variables $\hat{\lambda}_{1}^{\xi_{1}}$
and $\hat{\lambda}_{2}^{\xi_{1},\xi_{2}}$ do not depend on $\xi_{3}$.
So, in order to evaluate (\ref{eq:target}), the function $\phi$
needs to be evaluated only $\left(N_{T}\right)^{2}$ times in place
of $\left(N_{T}\right)^{3}$. This is a relevant improvement, because
evaluating the function $\phi$ many times can be time demanding.
Moreover, if the function $\phi$ is known only on the grid $\mathcal{G}$
-- this is what happens for the functions $\mathcal{\bar{V}}^{-}$
and $\mathcal{\bar{U}}^{-}$ -- then a two-dimensional interpolation
is required.

\subsection{Improving computational efficiency}

The Markov chain $\bar{Y}^{\Delta t}$ introduced in Appendix \ref{App0}
does not only provide a way to define the set $\mathcal{G}_{Y}$ but
it can be used to improve the evaluation of (\ref{eq:target}). Suppose
now $y$ in (\ref{eq:target}) to be equal to $y_{m}\in\mathcal{G}_{Y}$
for a particular integer $m$. Let $n=N_{T}\cdot i$ so that $\bar{t}_{n}=t_{i}$
and $\bar{t}_{n+N_{T}}=t_{i+1}$. In order to improve the discretisation
of the random variable $Y_{t_{i+1}}\left|Y_{t_{i}}=y_{m}\right.,$
we replace $\hat{\Lambda}_{1}^{N_{T}}$ in (\ref{eq:813}) with $\bar{Y}_{n+N_{T}}^{\Delta t}\left|\bar{Y}_{n}^{\Delta t}=y_{m}\right.$.
The transition probabilities 
\begin{equation}
\bar{p}_{m,l}=P\left(\bar{Y}_{n+N_{T}}^{\Delta t}=y_{l}\left|\bar{Y}_{n}^{\Delta t}=y_{m}\right.\right)
\end{equation}
 for $m,l\in\left\{ 0,\dots,2N_{Y}\right\} $ can be obtained by computing
the $N_{T}$-power of transition matrix of $\bar{Y}^{\Delta t}$,
whose elements are determined according to Table \ref{tab:Transition-probabilities-for}.
Finally, we conclude by observing that the support of the random variable
$\bar{Y}_{n+N_{T}}^{\Delta t}\left|\bar{Y}_{n}^{\Delta t}=y_{m}\right.$
is a subset of $\mathcal{G}_{Y}$ for every value $y_{m}$ in $\mathcal{G}_{Y}$.
We stress out that the support of the random variable $\hat{\Lambda}_{1}^{N_{T}}$
has $N_{T}+1$ elements, while the support of $\bar{Y}_{n+N_{T}}^{\Delta t}\left|\bar{Y}_{n}^{\Delta t}=y_{m}\right.$
has at most $2N_{Y}+1$ elements. Numerical tests show that $2N_{Y}+1$
is usually smaller than $N_{T}+1$, so computational efficiency is
improved: for example, with respect to our tests in Section \ref{Sec6},
we have $N_{T}+1=51$ and $2N_{Y}+1=27$.

The Markov chain $\bar{Y}^{\Delta t}$ helps us to discretize the
process $Y$ but in order to compute (\ref{eq:87}) we also have to
simulate the whole random vector $\hat{\Lambda}^{N_{T}}$ in (\ref{eq:82}).
To do so, we have to compute the normal Gaussian increments associated
to the transitions of $\bar{Y}^{\Delta t}$. Let us define the discrete
random variable $\bar{G}^{N_{T}}$ as the standard score of $\bar{Y}_{n+N_{T}}^{\Delta t}\left|\bar{Y}_{n}^{\Delta t}=y_{m}\right.$,
that is 
\begin{equation}
\bar{G}^{N_{T}}=\frac{\bar{Y}_{n+N_{T}}^{\Delta t}-\mu_{Y}^{1}\left(y_{m}\right)}{\sigma_{Y}^{1}}.
\end{equation}
Since $\bar{Y}_{n+N_{T}}^{\Delta t}\left|\bar{Y}_{n}^{\Delta t}=y_{m}\right.$
matches the first two moment of the random Gaussian variable $Y_{t_{i+1}}\left|Y_{t_{i}}=y_{m}\right.$,
then $\bar{G}^{N_{T}}$ matches the first two moments of a standard
Gaussian variable and so it can be employed in place of $\hat{G}_{1}^{N_{T}}$.
Moreover, $\mu_{Y}^{1}\left(y_{m}\right)=\mu_{1}$ and $\left(\sigma_{Y}^{1}\right)^{2}=\Pi_{11}$.
Then, we define the vector $\bar{\Lambda}^{N_{T}}=\left(\bar{\Lambda}_{1}^{N_{T}},\bar{\Lambda}_{2}^{N_{T}},\bar{\Lambda}_{3}^{N_{T}}\right)^{\top}$given
by
\begin{equation}
\bar{\Lambda}^{N_{T}}=\mu+\Gamma\left(\bar{G}_{1}^{N_{T}},\hat{G}_{2}^{N_{T}},\hat{G}_{3}^{N_{t}}\right)^{\top},
\end{equation}
which converges to $\Lambda$ and in particular $\bar{\Lambda}_{1}^{N_{T}}=\bar{Y}_{n+N_{T}}^{\Delta t}\left|\bar{Y}_{n}^{\Delta t}=y_{m}\right.$.
Let $\xi_{1},\xi_{2},\xi_{3}$ be three integers such that $\xi_{1}$
is in $\left\{ 0,\dots2N_{Y}\right\} $ and $\xi_{2},\xi_{3}$ are
in $\left\{ 0,\dots,N_{T}\right\} $. Let$\left(\bar{\lambda}_{1}^{\xi_{1}},\bar{\lambda}_{2}^{\xi_{1},\xi_{2}},\bar{\lambda}_{3}^{\xi_{1},\xi_{2},\xi_{3}}\right)^{\top}$
be the vector defined by
\begin{equation}
\left(\bar{\lambda}_{1}^{\xi_{1}},\bar{\lambda}_{2}^{\xi_{1},\xi_{2}},\bar{\lambda}_{3}^{\xi_{1},\xi_{2},\xi_{3}}\right)^{\top}=\mu+\Gamma\left(\bar{g}^{\xi_{1}},\hat{g}^{\xi_{2}},\hat{g}^{\xi_{3}}\right)^{\top},
\end{equation}
where $\left\{ \bar{g}^{0},\dots\bar{g}^{2N_{Y}}\right\} $ is the
support of $\bar{G}^{N_{T}}$. Please, note that $\bar{\lambda}_{1}^{\xi_{1}}$
is equal to $y_{\xi_{1}}$ which is in $\mathcal{G}_{Y}$. Thus, we
obtain the following approximation of $E$, based on the Markov chain
$\bar{Y}^{\Delta t}$:
\begin{align}
\bar{E} & =\sum_{\xi_{1}=0}^{2N_{Y}}\bar{p}_{m,\xi_{1}}\sum_{\xi_{2}=0}^{N_{T}}\hat{p}^{\xi_{2}}\phi\left(y_{\xi_{1}},\exp\left(\bar{\lambda}_{2}^{\xi_{1},\xi_{2}}\right),g,h\right)\sum_{\xi_{3}=0}^{N_{T}}\hat{p}^{\xi_{3}}\exp\left(-\bar{\lambda}_{3}^{\xi_{1},\xi_{2},\xi_{3}}\right).\label{eq:E2}
\end{align}

We conclude by observing that equation (\ref{eq:E2}) has one important
advantage over equation (\ref{eq:87}), that improves computational
efficiency when the function $\phi$ is known only at the points of
$\mathcal{G}$. Specifically, the computation of (\ref{eq:87}) requires
a two-dimensional interpolation to evaluate the function $\phi$ outside
$\mathcal{G}$ while (\ref{eq:E2}) requires only a one-dimensional
interpolation because, as opposed to $\hat{\lambda}_{1}^{\xi_{1}}$,
$y_{\xi_{1}}$ is an element of $\mathcal{G}_{Y}$. To this aim, we
employ one-dimensional cubic spline interpolation, which is very fast
and accurate.

\FloatBarrier

\bibliographystyle{abbrv}
\bibliography{bibliography}

\begin{thebibliography}{10}

\bibitem{bacinello2019}
A.~R. Bacinello and I.~Zoccolan.
\newblock Variable annuities with a threshold fee: valuation, numerical
  implementation and comparative static analysis.
\newblock {\em Decisions in Economics and Finance}, 42(1):21--49, 2019.

\bibitem{bernard2016}
C.~Bernard and M.~Kwak.
\newblock Semi-static hedging of variable annuities.
\newblock {\em Insurance: Mathematics and Economics}, 67:173--186, 2016.

\bibitem{brigo2007}
D.~Brigo and F.~Mercurio.
\newblock {\em {Interest Rate Models-Theory and Practice: with Smile, Inflation
  and Credit}}.
\newblock Springer Science \& Business Media, 2007.

\bibitem{costabile2017}
M.~Costabile.
\newblock A lattice-based model to evaluate variable annuities with guaranteed
  minimum withdrawal benefits under a regime-switching model.
\newblock {\em Scandinavian Actuarial Journal}, 2017(3):231--244, 2017.

\bibitem{costabile2020}
M.~Costabile, I.~Massab\`o, and E.~Russo.
\newblock Evaluating variable annuities with {GMWB} when exogenous factors
  influence the policy-holder's withdrawals.
\newblock {\em The European Journal of Finance}, 26(2-3):238--257, 2020.

\bibitem{dai2015}
T.-S. Dai, S.~S. Yang, and L.-C. Liu.
\newblock Pricing guaranteed minimum/lifetime withdrawal benefits with various
  provisions under investment, interest rate and mortality risks.
\newblock {\em Insurance: Mathematics and Economics}, 64:364--379, 2015.

\bibitem{donnelly2012}
R.~F. Donnelly, S.~Jaimungal, and D.~Rubisov.
\newblock Valuing guaranteed withdrawal benefits with stochastic interest rates
  and volatility.
\newblock {\em Quantitative Finance}, 14(2):369--382, 2014.

\bibitem{ekvall1996}
N.~Ekvall.
\newblock A lattice approach for pricing of multivariate contingent claims.
\newblock {\em European Journal of Operational Research}, 91(2):214--228, 1996.

\bibitem{forsyth2014}
P.~Forsyth and K.~Vetzal.
\newblock An optimal stochastic control framework for determining the cost of
  hedging of variable annuities.
\newblock {\em Journal of Economic Dynamics and Control}, 44:29--53, 2014.

\bibitem{gomes2009}
A.~Gomes, I.~Voiculescu, J.~Jorge, B.~Wyvill, and C.~Galbraith.
\newblock {\em {Implicit Curves and Surfaces: Mathematics, Data Structures and
  Algorithms}}.
\newblock Springer Science \& Business Media, 2009.

\bibitem{goudenege2016}
L.~Gouden{\`e}ge, A.~Molent, and A.~Zanette.
\newblock Pricing and hedging {GLWB} in the {Heston} and in the
  {Black--Scholes} with stochastic interest rate models.
\newblock {\em Insurance: Mathematics and Economics}, 70:38--57, 2016.

\bibitem{goudenege2018}
L.~Gouden{\`e}ge, A.~Molent, and A.~Zanette.
\newblock Pricing and hedging {GMWB} in the {Heston} and in the
  {Black--Scholes} with stochastic interest rate models.
\newblock {\em Computational Management Science}, 16(1), 2018.

\bibitem{goudenege2019}
L.~Gouden{\`e}ge, A.~Molent, and A.~Zanette.
\newblock Gaussian process regression for pricing variable annuities with
  stochastic volatility and interest rate.
\newblock {\em arXiv preprint arXiv:1903.00369}, 2019.

\bibitem{gudkov2017}
N.~Gudkov, K.~Ignatieva, and J.~Ziveyi.
\newblock Pricing of guaranteed minimum withdrawal benefits in variable
  annuities under stochastic volatility, stochastic interest rates and
  stochastic mortality via the componentwise splitting method.
\newblock {\em Quantitative Finance}, 19(3):501--518, 2019.

\bibitem{haentjens2012}
T.~Haentjens and K.~J. In't~Hout.
\newblock Alternating direction implicit finite difference schemes for the
  {Heston-Hull-White} partial differential equation.
\newblock {\em The Journal of Computational Finance}, 16(1):83, 2012.

\bibitem{hull1994}
J.~Hull and A.~White.
\newblock Numerical procedures for implementing term structure models {I}:
  Single-factor models.
\newblock {\em Journal of derivatives}, 2(1):7--16, 1994.

\bibitem{lin2016}
X.~S. Lin, P.~Wu, and X.~Wang.
\newblock Move-based hedging of variable annuities: A semi-analytic approach.
\newblock {\em Insurance: Mathematics and Economics}, 71:40--49, 2016.

\bibitem{mackay2017}
A.~MacKay, M.~Augustyniak, C.~Bernard, and M.~R. Hardy.
\newblock Risk management of policyholder behavior in equity-linked life
  insurance.
\newblock {\em Journal of Risk and Insurance}, 84(2):661--690, 2017.

\bibitem{moenig2016}
T.~Moenig and D.~Bauer.
\newblock Revisiting the risk-neutral approach to optimal policyholder
  behavior: A study of withdrawal guarantees in {Variable Annuities}.
\newblock {\em Review of Finance}, 20(2):759--794, 2016.

\bibitem{moenig2018}
T.~Moenig and N.~Zhu.
\newblock Lapse-and-reentry in {Variable Annuities}.
\newblock {\em Journal of Risk and Insurance}, 85(4):911--938, 2018.

\bibitem{winsconsin2017}
S.~Moran.
\newblock Taxation of insurance companies.
\newblock \url{https://docs.legis.wisconsin.gov/misc/lfb/informational_papers},
  2017.
\newblock Wisconsin Legislative Fiscal Bureau.

\bibitem{nelson1990}
D.~B. Nelson and K.~Ramaswamy.
\newblock Simple binomial processes as diffusion approximations in financial
  models.
\newblock {\em The review of financial studies}, 3(3):393--430, 1990.

\bibitem{nissim2010}
D.~Nissim.
\newblock Analysis and valuation of insurance companies.
\newblock {\em Center for Excellence in Accounting and Security Analysis}, (2),
  2010.

\bibitem{peng2012}
J.~Peng, K.~S. Leung, and Y.~K. Kwok.
\newblock Pricing guaranteed minimum withdrawal benefits under stochastic
  interest rates.
\newblock {\em Quantitative Finance}, 12(6):933--941, 2012.

\bibitem{pollard2012convergence}
D.~Pollard.
\newblock {\em Convergence of Stochastic Processes}.
\newblock Springer Science \& Business Media, 2012.

\bibitem{ross1987}
S.~Ross.
\newblock Arbitrage and martingales with taxation.
\newblock {\em Journal of Political Economy}, 95(2):371--393, 1987.

\bibitem{SRI2019}
{Secure Retirement Institute}.
\newblock {U.S.} individual annuity sales survey, third quarter 2019.
\newblock
  \url{https://www.limra.com/globalassets/limra/newsroom/fact-tank/sales-data/2019/q4/4q-2019-annuity-sales-estimates-vfinal.pdf}.
\newblock Accessed: 01 Apr 2020.

\bibitem{shevchenko2017}
P.~V. Shevchenko and X.~Luo.
\newblock Valuation of variable annuities with guaranteed minimum withdrawal
  benefit under stochastic interest rate.
\newblock {\em Insurance: Mathematics and Economics}, 76:104--117, 2017.

\bibitem{skipper2001}
H.~D. Skipper~Jr.
\newblock The taxation of life insurance policies in {OECD} countries:
  Implications for tax policy and planning.
\newblock {\em Insurance and Private Pensions Compendium for Emerging
  Economies. Paris: OECD}, 2001.

\bibitem{Table}
{Social Security Administration}.
\newblock {Actuarial Life Table}.
\newblock \url{https://www.ssa.gov/oact/STATS/table4c6_2007.html}.

\end{thebibliography}

\end{document}